\documentclass[acmsmall,screen,nonacm]{acmart}

\usepackage{subcaption}
\usepackage{xspace}
\usepackage{setspace}
\usepackage{pifont}
\usepackage{enumitem}
\usepackage{microtype}
\usepackage[separate-uncertainty]{siunitx}
\usepackage{fancybox}
\usepackage{tikz}

\usepackage{fnpct} 

\usepackage{hyperref}
\usepackage{cleveref}

\setcopyright{acmcopyright}
\copyrightyear{2018}
\acmYear{2018}
\acmDOI{XXXXXXX.XXXXXXX}

\acmPrice{15.00}
\acmISBN{978-1-4503-XXXX-X/18/06}

\newcommand{\pqplan}{{\normalfont\textsc{Syn}\-\textsc{chro}\-\textsc{nized} \textsc{Pla}\-\textsc{nari}\-\textsc{ty}}\xspace}
\newcommand{\cplan}{{\normalfont\textsc{Clus}\-\textsc{tered} \textsc{Pla}\-\textsc{nari}\-\textsc{ty}}\xspace}

\newcommand{\sefe}{{\normalfont\textsc{SEFE}}\xspace}
\newcommand{\consefe}{{\normalfont\textsc{Con}\-\textsc{nect}\-\textsc{ed} \textsc{SEFE}}\xspace}

\newcommand{\lplan}{{\normalfont\textsc{Level} \textsc{Pla}\-\textsc{nari}\-\textsc{ty}}\xspace}

\newcommand{\pTcTwoPage}{{\normalfont\textsc{Par}\-\textsc{ti}\-\textsc{tioned} $\mathcal{T}$-\textsc{coherent} 2-\textsc{page} \textsc{Book} \textsc{Em}\-\textsc{bed}\-\textsc{ding}}\xspace}
\newcommand{\pTwoPage}{{\normalfont\textsc{Par}\-\textsc{ti}\-\textsc{tioned} 2-\textsc{page} \textsc{Book} \textsc{Em}\-\textsc{bed}\-\textsc{ding}}\xspace}

\newcommand{\contract}{\texttt{En}\-\texttt{cap}\-\texttt{su}\-\texttt{late}\-\texttt{And}\-\texttt{Join}\xspace}
\newcommand{\propagate}{\texttt{Prop}\-\texttt{a}\-\texttt{gatePQ}\xspace}
\newcommand{\simplify}{\texttt{Sim}\-\texttt{pli}\-\texttt{fy}\-\texttt{Match}\-\texttt{ing}\xspace}

\newcommand{\SP}{\texttt{SP}\xspace}
\newcommand{\ILP}{\texttt{ILP}\xspace}
\newcommand{\HT}{\texttt{HT}\xspace}
\newcommand{\HTf}{\texttt{HT-f}\xspace}

\newcommand{\combflag}[1]{[\texttt{#1}]}
\newcommand{\combconf}[1]{{\SP}\texttt{[#1]}}
\newcommand{\SPd}{\combconf{d}\xspace}

\newcommand{\dsold}{\texttt{C-OLD}\xspace}
\newcommand{\dsmed}{\texttt{C-MED}\xspace}
\newcommand{\dsmedncp}{\texttt{C-NCP}\xspace}
\newcommand{\dslarge}{\texttt{C-LRG}\xspace}
\newcommand{\dspq}{\texttt{SP-LRG}\xspace}
\newcommand{\dssefe}{\texttt{SEFE-LRG}\xspace}

\newcommand{\dsoldsize}{\num{1643}}
\newcommand{\dsmedncpsize}{\num{13834}}
\newcommand{\dsmedsize}{\num{5171}}
\newcommand{\dslargesize}{\num{5096}}
\newcommand{\dssefesize}{\num{1008}}
\newcommand{\dspqsize}{\num{1587}}
\newcommand{\dstotalsize}{\num{28339}}

\begin{document}

\title{Constrained Planarity in Practice}
\subtitle{Engineering the Synchronized Planarity Algorithm}

\author{Simon D. Fink}
\email{finksim@fim.uni-passau.de}
\orcid{0000-0002-2754-1195}
\author{Ignaz Rutter}
\email{rutter@fim.uni-passau.de}
\orcid{0000-0002-3794-4406}
\affiliation{\institution{University of Passau}
  \department{Faculty of Computer Science and Mathematics}
  \city{Passau}
  \country{Germany}
}

\begin{abstract}
In the constrained planarity setting, we ask whether a graph admits a planar drawing that additionally satisfies a given set of constraints.
  These constraints are often derived from very natural problems;
  prominent examples are \lplan, where vertices have to lie on given horizontal lines indicating a hierarchy, and
  \cplan, where we additionally draw the boundaries of clusters which recursively group the vertices in a crossing-free manner.
Despite receiving significant amount of attention and substantial theoretical progress on these problems, 
  only very few of the found solutions have been put into practice and evaluated experimentally.\\
  In this paper, we describe our implementation of the recent quadratic-time algorithm by Bläsius et al.~\cite{bfr-spw-21} for solving the problem \pqplan,
  which can be seen as a common generalization of several constrained planarity problems, including the aforementioned ones.
  Our experimental evaluation on an existing benchmark set shows that even our baseline implementation outperforms all competitors by at least an order of magnitude.
  We systematically investigate the degrees of freedom in the implementation of the \pqplan algorithm for larger instances and propose several modifications that further improve the performance.
  Altogether, this allows us to solve instances with up to 100 vertices in milliseconds and instances with up to \num{100000} vertices within a few minutes.
\end{abstract}

\maketitle

\begin{acks}
Funded by the Deutsche Forschungsgemeinschaft (German Research Foundation, DFG) under grant RU-1903/3-1.
\end{acks}

\section{Introduction}
In many practical graph drawing applications we not only seek any drawing that maximizes legibility, but also want to encode additional information via certain aspects of the underlying layout.
Examples are \emph{hierarchical} drawings like organizational charts, where we encode a hierarchy among vertices by placing them on predefined levels,
\emph{clustered} drawings, where we group vertices by enclosing them in a common region, and
\emph{animated} drawings, where changes to a graph are shown in steps while keeping a static part fixed.
In practice, clustered drawings are for example UML diagrams, where classes are grouped according to the package they are contained in, computer networks, where devices are grouped according to their subnetwork, and integrated circuits, where certain components should be placed close to each other.
As crossings negatively affect the readability of drawings~\cite{pac-gla-02,wpcm-cmo-02}, we preferably seek planar, i.e.\ crossing-free, drawings.
The combination of these concepts leads to the field of constrained planarity problems, where we ask whether a graph admits a planar drawing that satisfies a given set of constraints.
This includes the problems \lplan \cite{jlm-lpt-98,brue-pvf-21}, \cplan \cite{br-anp-16,len-hpt-89,fce-pfc-95}, and \textsc{Simultaneous Embedding with Fixed Edges (SEFE)} \cite{bcd-osp-07,bkr-seo-13,rut-se-20}, which respectively model the aforementioned applications; see \Cref{fig:constplan-examples}.
Formally, these problems are defined as follows.

\begin{center}
  \includegraphics[page=1]{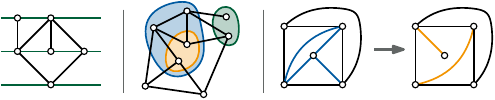}
  \captionof{figure}{
    Examples of constrained planarity problems: \lplan (a), \cplan (b), \sefe~(c).
  }
  \label{fig:constplan-examples}
\end{center}

\cornersize*{10pt}
\setlength\fboxsep{10pt}

\newcommand{\problem}[3]{\Ovalbox{\begin{minipage}{.9\textwidth}\textbf{Problem} #1
  \begin{description}[nosep,topsep=.2cm]
    \item[Given] #2
    \item[Question] #3
  \end{description}
\end{minipage}}}

\newcommand*\circled[1]{ \protect\tikz[baseline=(char.base)]{ \protect\node[shape=circle,draw,inner sep=0.2pt,scale=0.7] (char) {#1};}}

\problem{\lplan}{
  graph $G$, leveling function $\ell : V(G) \to \mathbb{N}$
}{
  Is there a planar drawing where each vertex $v\in V(G)$ has y-coordinate $\ell(v)$
  and all edges are drawn y-monotone?
}

\problem{\cplan}{
  graph $G$, rooted cluster tree $T$, cluster assignment function $\gamma : V(G) \to V(T)$
}{
  Is there a planar drawing where, for each cluster $c\in V(T)$, we can add a simple closed region that
  \begin{enumerate}[leftmargin=0pt,nosep]
  \item encloses exactly the vertices mapped to $c$ or one of its descendants in $T$, and
  \item has a border that crosses each edge that connects a vertex within its interior to a vertex on its outside exactly once, but no other edge or cluster region border?
  \end{enumerate}
}

\problem{\sefe}{
  graphs $G^{\circled{1}}, G^{\circled{2}}$ with a shared graph $G = G^{\circled{1}} \cap G^{\circled{2}}$
}{
  Are there a planar drawings of $G^{\circled{1}}$ and $G^{\circled{2}}$ that induce the same drawing of their shared part $G$?
}\\

In the last years, the family formed by these problems and other variants of constrained planarity received a lot of attention in the field of Graph Drawing.
Efficient algorithms were discovered for many of them, while a few others turned out to be NP-complete; see \cite{sch-tat-13} and \cite{dloz-pgw-15} for an overview.
In contrast to the extensive theoretical considerations and the direct motivation by applications, only very few of the found algorithms (many of which have a linear or at most quadratic asymptotic running time) have been implemented and evaluated in practice.
This also contrasts the wide variety of implementations available for the different linear-time algorithms for ordinary, i.e., unconstrained planarity~\cite{pat-pta-13}, which have also been thoroughly assessed in terms of their practical running time~\cite{fmr-tta-06,bcpb-smy-04}.

The recently introduced problem \pqplan~\cite{bfr-spw-21} not only generalizes many constrained planarity variants, among them in particular \textsc{Level} and \cplan as well as variants of \textsc{SEFE}, but also has a comparatively simple quadratic-time solution.
Akin to the Goldberg and Tarjan push-relabel algorithm~\cite{gt-ana-88}, it uses few and simple operations that can be applied in arbitrary order.
Through reductions from many other problems (see~\Cref{fig:constplan-schemaequivalence} for an overview), an implementation would also allow to solve other constrained planarity problems for which no practical solution is available.
This wide area of possible applications and the fact that the algorithm offers several degrees of freedom make it an ideal starting point for algorithm engineering.

In this paper, we describe our implementation of the \pqplan algorithm, which we evaluate by comparing its results and running times to those of two existing implementations for the \cplan problem.
We complement the previous theoretical running time analysis by Bläsius et al.~\cite{bfr-spw-21} with practical measurements, highlighting which parts of the algorithm take the most time.
Based on this, we engineer the algorithm by analyzing how to best employ the degrees of freedom present in the algorithm and by proposing algorithmic improvements to overcome performance bottlenecks.
\Cref{sec:syncplan} provides more background on constrained planarity and \pqplan in particular, while \Cref{sec:related-work} gives an overview of previous practical approaches to constrained planarity.
In \Cref{sec:eval-old} we describe our implementation of \pqplan and evaluate its performance in comparison with the two other available \cplan implementations.
We tune the running time of our implementation to make it practical even on large instances in \Cref{sec:engineering}.
We analyze the effects of our engineering in greater detail in \Cref{sec:further-analysis}.

\begin{figure}[t]
  \centering
  \includegraphics[page=1]{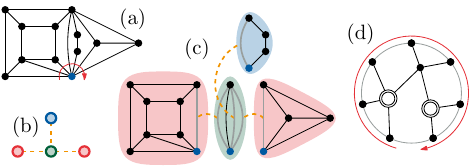}
  \caption{
    A planar graph (a), its SPQR-tree (b) and the corresponding skeletons (c).
    Rigids are highlighted in red, parallels in green, and series in blue.
    The embedding tree of the vertex marked in blue (d).
    Small black disks are P-nodes, larger white disks are Q-nodes.
  }
  \label{fig:spqrtree}
\end{figure}

\subsection{Preliminaries.}
We rely on some well-known concepts from the fields of graph drawing and planar graphs.
We only briefly define the most important terms here and refer to the theoretical description of the implemented algorithm \cite{bfr-spw-21} for more comprehensive definitions.
A more gentle introduction to the concepts can also be found in Chapter~1 of the Handbook of Graph Drawing and Visualization \cite{pat-pta-13}.
We consider two planar (i.e., crossing-free) drawings equivalent if they define the same \emph{rotation system}, which specifies for each vertex its \emph{rotation}, i.e., the cyclic order of the edges around the vertex.
An \emph{embedding} is an equivalence class of planar drawings induced by this relation.
An \emph{embedding tree}~\cite{bfr-spw-21} is a PQ-tree~\cite{bl-tft-76} that describes all possible rotations of a vertex in a planar graph; see \Cref{fig:spqrtree}d.
Its leaves correspond to the incident edges, while its inner nodes are either Q-nodes, which dictate a fixed ordering of their incident subtrees that can only be reversed, or are P-nodes, which allow arbitrary permutation.
A BC-tree describes the decomposition of a connected graph into its \emph{biconnected} components, which cannot be disconnected by the removal of a so-called \emph{cut-vertex}.
Each node of a BC-tree represents either a cut-vertex or a maximal biconnected \emph{block}.
We refer to a vertex that is not a cut-vertex as \emph{block-vertex}.
An SPQR-tree~\cite{dbt-olm-96} describes the decomposition of a biconnected graph into its \emph{triconnected} components, which cannot be disconnected by the removal of a so-called \emph{split-pair} of two vertices.
Each inner node represents a \emph{skeleton}, which is either a triconnected \emph{`rigid'} minor whose planar embedding can only be mirrored, a split-pair of two \emph{pole} vertices connected by multiple \emph{`parallel'} subgraphs that can be permuted arbitrarily, or a cycle formed by split-pairs separating a \emph{`series'} of subgraphs; see \Cref{fig:spqrtree}c.
All three kinds of trees can be computed in time linear in the size of the given graph~\cite{br-spo-16,pat-pta-13,gm-alt-01}.

\section{Constrained Planarity}\label{sec:syncplan}

\begin{figure}[t]
  \centering
  \includegraphics[page=4,width=.75\textwidth]{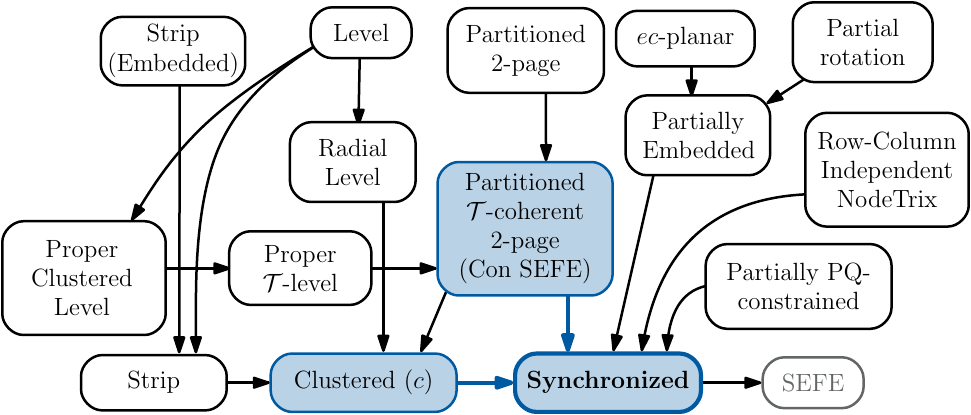}
  \caption{Constrained planarity variants related to \pqplan, updated selection from~\cite{dloz-pgw-15}. Problems and reductions marked in blue are used for generating test instances.}
  \label{fig:constplan-schemaequivalence}
\end{figure}

Schaefer~\cite[Figure 2]{sch-tat-13} introduced a hierarchy on the various variants of constrained planarity that have been studied in the past.
\Cref{fig:constplan-schemaequivalence} shows a subset of this hierarchy, incorporating updates up to 2015 by Da Lozzo~\cite[Figure 0.1]{dloz-pgw-15}.
Arrows indicate that the target problem either generalizes the source problem or solves it via a reduction.
In the version of Da Lozzo, the problems \textsc{Strip, Clustered} and \pqplan as well as (\textsc{Connected}) \textsc{SEFE} still formed a frontier of problems with unknown complexity, separating efficiently solvable problems from those that are NP-hard.
Since then many of these problems were settled in P, especially due to the \cplan solution from 2019 by Fulek and Tóth~\cite{ft-aec-22}.
The only problem from this hierarchy that remains with an unknown complexity is \sefe.
In this section, we want to give a short summary of the history of \cplan and \sefe,
which we see central to the field of constrained planarity and which also serve as a motivation for \pqplan.
Afterwards, we will give a short summary of the algorithm we implement for solving the latter problem.
We point the interested reader to the original description \cite{bfr-spw-21} for full details.

Recall that in \sefe, we are given two graphs that share some common part and we want to embed both graphs individually such that their common parts are embedded the same way~\cite{bcd-osp-07,bkr-seo-13,rut-se-20}.
More general \sefe variants are often NP-complete, e.g., the case with three given graphs~\cite{gjp-sge-06}, even if all share the same common part~\cite{aln-aos-15,sch-tat-13}.
In contrast, more restricted variants are often efficiently solvable, e.g., when the shared graph is biconnected, a star, a set of cycles, or has a fixed embedding~\cite{abf-tts-12,br-dar-15,abf-tpo-15}.
The case where the shared graph is connected, which is called \consefe, was shown to be equivalent to the so-called \pTcTwoPage problem \cite{abf-tts-12} and to be reducable to \cplan~\cite{al-sc-16}, all of which were recently shown to be efficiently solvable~\cite{ft-aec-22}.
In contrast to these results, the complexity of the general \sefe problem with two graphs sharing an arbitrary common graph is still unknown.

\begin{figure}
    \includegraphics{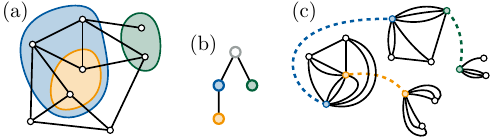}
    \caption{A \cplan instance (a), its cluster tree (b), and its CD-tree representation (c).}
    \label{fig:cdtree}
\end{figure}

Recall that in \cplan, the embedding has to respect a laminar family of clusters, that is every vertex is assigned to some (hierarchically nested) cluster and an edge may only cross a the border of a cluster's region if it connects a vertex from the inside with one from the outside~\cite{br-anp-16,len-hpt-89}; see \Cref{fig:cdtree} for an example.
Lengauer~\cite{len-hpt-89} studied and solved this problem as early as 1989 in the setting where the clusters are connected. Feng et al.~\cite{fce-pfc-95}, who coined the term \cplan, rediscovered this algorithm and asked the general question where disconnected clusters are allowed.  This question remained open for 25 years.
In that time, polynomial-time algorithms were found for many special-cases~\cite{adl-cpw-19,cdbf-cpo-08,fkmp-cpt-15,gjl-aic-02} before Fulek and Tóth~\cite{ft-aec-22} found an $O((n+d)^8)$ solution in 2019, where $d$ is the number of crossings between a cluster-border and an edge leaving the cluster.
Shortly thereafter, Bläsius et al.~\cite{bfr-spw-21} gave a solution with running time in $O((n+d)^2)$ that works via a linear-time reduction to \pqplan. 

In \pqplan, we are given a graph together with a set of \emph{pipes}, each of which pairs up two distinct vertices of the graph.
Each pipe synchronizes the rotation of its two paired-up vertices (its \emph{endpoints}) in the following sense:
We seek a planar embedding of the graph where for each pipe $\rho$, the rotations of its endpoints line up under the bijection $\varphi_\rho$ associated with~$\rho$~\cite{bfr-spw-21}.
Formally, this problem is defined as follows.\\

\problem{\pqplan\footnote{Note that we disregard the originally included Q-vertices here, as they can also be modeled using pipes~\cite[Section 5]{bfr-spw-21}.}}{
  graph $G$ and a set $\mathcal P$, where each \emph{pipe} $\rho\in\mathcal P$ consists of two distinct vertices $v_1, v_2\in V(G)$ and a bijection $\varphi_\rho$ between the edges incident to $v_1$ and those incident to $v_2$, and each vertex is part of at most one pipe
}{
  Is there a drawing of $G$ where for each pipe $\rho=(v_1,v_2,\varphi_\rho)$, the cyclic order of edges incident to $v_1$ lines up with the order of edges incident to $v_2$ under the bijection $\varphi_\rho$?
}\\

The motivation for this ``synchronization'' can best be seen by considering the reduction from \textsc{Clustered} to \pqplan.
At each cluster boundary, we split the graph into two halves: one where we contract the inside of the cluster into a single vertex and one where we contract the outside into a single vertex.
In a clustered planar embedding, the order of the edges ``leaving'' one cluster (i.e. the rotation of its contracted vertex in the one half) needs to match the order in which they ``enter'' the parent cluster (i.e. the the rotation of the corresponding contracted vertex in the other half).
The graph resulting from separately contracting each side of a cluster boundary is called CD-tree~\cite{br-anp-16}; see~\Cref{fig:cdtree} and~\cite[Figure 6]{bfr-spw-21} for an example.
Using this graph, the synchronization of rotations can easily be modeled via \pqplan by pairing the two contracted vertices corresponding to the same cluster boundary with a pipe.

\begin{figure}[t]
  \centering
  \includegraphics[page=1,scale=1]{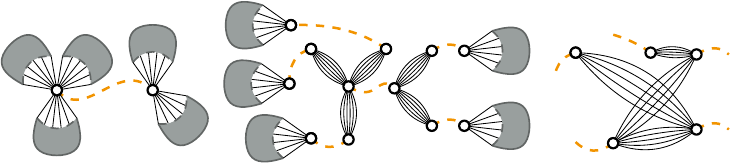}\\[.25cm]
  \includegraphics[page=2,scale=1]{graphics/syncplan}\\[.2cm]
  \includegraphics[page=3,scale=1]{graphics/syncplan}
  \caption{
    The operations for solving \pqplan~\cite{bfr-spw-21}.
    Pipes are indicated by orange dashed lines, their endpoints are shown as larger disks.
\textbf{Top:}
      Two cut-vertices paired-up by a pipe (left), the result of encapsulating their incident blocks (middle) and the bipartite graph resulting from joining both cut-vertices (right).
    \textbf{Middle:}
      A block-vertex pipe endpoint (left) that has a non-trivial embedding tree (middle) that is propagated to replace both the vertex and its partner (right).
    \textbf{Bottom:}
      Three different cases of paired-up vertices with trivial embedding trees (blue) and how their pipes can be removed or replaced (red).
}
  \label{fig:syncplan-ops}
\end{figure}

In the quadratic algorithm for solving \pqplan, a pipe is \emph{feasible} if one of the three following operation can be applied to remove it.
\begin{description}
\item[\contract]
If both endpoints of the pipe are cut-vertices, they are ``encapsulated'' by collapsing each incident block to a single vertex to obtain two stars with paired-up centers.
Additionally, we split the original components at the two cut-vertices, so that each of their incident blocks is retained as separate component with its own copy of the cut-vertex.
These copies are synchronized with the respective vertex incident to the original cut-vertex representing the collapsed block.
Now the cut-vertices can be removed by ``joining'' both stars at their centers, i.e, by identifying their incident edges according to the given bijection; see the top row of \Cref{fig:syncplan-ops}.
\item[\propagate]
If one endpoint of the pipe is a block-vertex and has an embedding tree that not only consists of a single P-node (i.e., it is \emph{non-trivial}), a copy of this embedding tree is inserted (``propagated'') in place of each respective pipe endpoint.
The inner nodes of the embedding trees are synchronized by pairing corresponding vertices with a pipe; see the middle row of \Cref{fig:syncplan-ops}.
Note that, as Q-nodes only have a binary embedding decision, they can more easily be synchronized via a 2-SAT formula instead of using pipes.
\item[\simplify]
In the remaining case, at least one of the endpoints of the pipe is a block-vertex but has a trivial embedding tree.
If the vertex (or, more precisely, the parallel in the SPQR-tree that completely defines its rotation) can respect arbitrary rotations, we can simply remove the pipe.
When the other pole of the parallel is also paired-up and has a trivial embedding tree, we ``short-circuit'' the pipe across the parallel; see the bottom row of \Cref{fig:syncplan-ops}.
One exception is if the pipe matches the poles of the same parallel, where we can again remove the pipe without replacement.
\end{description}
The algorithm then works by simply applying a suitable operation on an \emph{arbitrary} feasible pipe each step.
Moreover, it can be shown that if a pipe is not feasible, then this is directly caused by a close-by pipe with endpoints of higher degree~\cite{bfr-spw-21}.
Especially, this means that maximum-degree pipes are always feasible.

Each of the three operations runs in time linear in the degree of the removed pipe once the embedding trees it depends on has been computed.
This is dominated by the time spent on computing the embedding tree, which is linear in the size of the considered biconnected component.
Every applied operation removes a pipe, but potentially introduces new pipes of smaller degree.
Bläsius et al.~\cite{bfr-spw-21} show that the progress made by the removal of a pipe always dominates the overhead of the newly-introduced pipes and that the number of operations needed to remove all pipes is limited by the total degree of all paired-up vertices.
Furthermore, the resulting instance without pipes can be solved and embedded in linear time.
An embedding of the input graph can then be obtained by undoing all changes made to the graph in reverse order while maintaining the found embedding.
The algorithm thus runs in the following three simple phases:
\enlargethispage{\baselineskip}
\begin{enumerate}
\item While pipes are left, choose and remove an arbitrary feasible pipe by applying an operation. \item Solve and embed the resulting pipe-free (\emph{reduced}) instance.
\item Undo all applied operations while maintaining the embedding.
\end{enumerate}

\section{Related Work}\label{sec:related-work}
Surprisingly, in contrast to their intense theoretical consideration, constrained planarity problems have only received little practical attention so far.
Of all variants, practical approaches to \cplan were studied the most, although all implementations predate the first fully-correct polynomial-time solution and thus either have an exponential worst-case running time or cannot solve all instances.
Chimani et al.~\cite{cgj-cmc-08} studied the problem of finding maximal cluster planar subgraphs in practice using an Integer Linear Program (ILP) together with a branch-and-cut algorithm.
A later work~\cite{ck-sts-12} strengthened the ILP for the special case of testing \cplan, further improving the practical running time.
The work by Gutwenger et al.~\cite{gms-pew-14} takes a different approach by using a Hanani-Tutte-style formulation of the problem based on the work by Schaefer~\cite{sch-tat-13}.
Unfortunately, their polynomial-time testing algorithm cannot solve all instances and declines to make a decision for some instances.
The Hanani-Tutte-approach solved instances with up to 60 vertices and 8 clusters in up to half a minute, while the ILP approach only solves roughly \qty{90}{\percent} of these instances within 10 minutes~\cite{gms-pew-14}.

The only other constrained planarity variant for which we could find experimental results is \pTwoPage.
Angelini et al.~\cite{abb-iap-12} describe an implementation of the SPQR-tree-based linear-time algorithm by Hong and Nagamochi~\cite{hn-tpb-09}, which solves instances with up to \num{100000} vertices and two clusters in up to 40 seconds.
Unfortunately, their implementation is not publicly available.
For (\textsc{Radial}) \lplan, prototypical implementations were described in the dissertations by Leipert~\cite{lei-lpt-98} and Bachmaier~\cite{bac-cpo-04},
although in both cases neither further information, experimental results, nor source code is available.
The lack of an accessible and correct linear-time implementation may be due to the high complexity of the linear-time algorithms~\cite{brue-pvf-21}.
Simpler algorithms with a super-linear running time have been proposed~\cite{hh-plp-07,rsb-asf-01,fpss-hmd-12}.
For these, we could only find an implementation by Estrella-Balderrama et al.~\cite{efk-gat-10} for the quadratic algorithm by Harrigan and Healy~\cite{hh-plp-07}.
Unfortunately, this implementation has not been evaluated experimentally and we were also unable to make it usable independently of its Microsoft Foundation Classes GUI, with which it is tightly intertwined.

We are not aware of further practical approaches for constrained planarity variants.
Note that while the problems \pTwoPage and \lplan have linear-time solutions, they are much more restricted than \pqplan (see \Cref{fig:constplan-schemaequivalence}) and have no usable implementations available.
We thus focus our comparison on solutions to the \cplan problem which, besides being a common generalization of both other problems, fortunately also has all relevant implementations available.

\begin{table}[t]
  \centering
  \resizebox*{\linewidth}{!}{
  \sisetup{mode = math, range-phrase = --, table-align-text-before = true, table-align-comparator = false, table-align-text-after = false}
  \begin{tabular}{
      l | S[table-format=5] |
      r S[table-format={(}5.1{)}] |l|
      S[table-format={<=}3, table-align-text-before = false] S[table-format={(}2.1{)}] |
      S[table-format={<=}5, table-align-text-before = false] S[table-format={(}4.1{)}] |
      S[table-format={<=}7, table-align-text-before = false] S[table-format={(}5.1{)}]
  }
    Dataset   & {\#}  & \multicolumn{2}{c|}{Vertices} & {Density}    & \multicolumn{2}{c|}{Components} & \multicolumn{2}{c|}{Clusters/Pipes} & \multicolumn{2}{c}{$d$} \\\hline\rule{0pt}{2.2ex}\dsold    & 1643  & $\leq$$59$   & {(}17.2{)}    & \numrange{0.9}{2.2} (1.4) & =1      &            & \leq 19    & {(}4.2{)}    & \leq 256     & {(}34.0{)}    \\
    \dsmedncp & 13834 & $\leq$$500$  & {(}236.8{)}   & \numrange{0.6}{2.9} (1.9) & \leq48  & {(}21.7{)} & \leq 50    & {(}16.8{)}   & \leq 5390    & {(}783.3{)}   \\
    \dsmed    & 5171  & $\leq$$10^3$ & {(}311.6{)}   & \numrange{0.9}{2.9} (2.3) & \leq10  & {(}5.1{)}  & \leq 53    & {(}16.1{)}   & \leq 7221    & {(}831.8{)}   \\\hline\rule{0pt}{2.2ex}\dslarge  & 5096  & $\leq$$10^5$ & {(}15214.1{)} & \numrange{0.5}{3.0} (2.4) & \leq100 & {(}29.8{)} & \leq 989   & {(}98.8{)}   & \leq 2380013 & {(}44788.7{)} \\
    \dssefe   & 1008  & $\leq$$10^4$ & {(}3800.0{)}  & \numrange{1.1}{2.4} (1.7) & =1      &            & \leq 20000 & {(}7600.0{)} & \leq 113608  & {(}34762.4{)} \\
    \dspq     & 1587  & $\leq$$10^5$ & {(}25496.6{)} & \numrange{1.3}{2.5} (2.0) & \leq100 & {(}34.5{)} & \leq 20000 & {(}1467.4{)} & \leq 139883  & {(}9627.5{)}  \\
  \end{tabular}
  }
  \caption{Statistics for our different datasets, values in parentheses are averages. Column \# shows the number of instances while column $d$ shows the total number of cluster-border edge crossings or the total degree of all pipes, depending on the underlying instances.}
  \label{tab:datasets}
\end{table}

\section{Clustered Planarity in Practice}\label{sec:eval-old}
In this section, we shortly describe our C++ implementation of the \pqplan algorithm by Bläsius et al.~\cite{bfr-spw-21} and compare its running time and results on instances derived from \cplan with those of the two existing implementations by Chimani et al.~\cite{cgj-cmc-08,ck-sts-12} and by Gutwenger et al.~\cite{gms-pew-14}.
We base our implementation on the graph data structures provided by the OGDF~\cite{cgj-tog-13} and, as the only other dependency, use the PC-tree implementation by Fink et al.~\cite{fpr-eco-21} for the embedding trees.
The PC-tree is a datastructure that is conceptually equivalent to the PQ-tree we use as embedding tree, but is faster in practice~\cite{fpr-eco-21}.

The algorithm for \pqplan makes no restriction on how the next feasible pipe should be chosen.
For now, we will use a heap to always use a pipe of maximum degree, as this ensures that the pipe is feasible.
The operations used for solving \pqplan heavily rely on (bi-)connectivity information while also making changes to the graph that may affect this information.
As recomputing the information before each step would pose a high overhead, we maintain this information in the form of a BC-forest (i.e. a collection of BC-trees).
To generate the embedding trees needed by the \propagate and \simplify operations, we implement the Booth-Lueker algorithm for testing planarity~\cite{bl-tft-76,pat-pta-13} using PC-trees.
We use that, after processing all vertices of a biconnected component, the resulting PC-tree corresponds to the embedding tree of the vertex that was processed last.

\subsection{Evaluation Set-Up.}
We compare our implementation of \pqplan with the \cplan implementations \ILP by Chimani et al.~\cite{cgj-cmc-08,ck-sts-12} and \HT by Gutwenger at al.~\cite{gms-pew-14}.
Both are written in C++ and are part of the OGDF.
The \ILP implementation by Chimani et al.~\cite{cgj-cmc-08,ck-sts-12} uses the ABACUS ILP solver~\cite{ABACUS} provided with the OGDF.
We refer to our \pqplan implementation processing pipes in descending order of their degree as \SPd.
We use the embedding it generates for yes-instances as certificate to validate all positive answers.
For the Hanani-Tutte algorithm, we give the running times for the modes with embedding generation and verification (\HT) and the one without (\HTf) separately.
Note that \HTf only checks an important necessary, but not sufficient condition and thus may falsely classify negative instances as positive, see \cite[Figure 3]{gms-pew-14} and \cite[Figure 16]{fkmp-cpt-15} for examples where this is the case.
Variant \HT tries to verify a positive answer by generating an embedding, which works by incrementally fixing parts of a partial embedding and subsequently re-running the test.
This process may fail at any point, in which case the algorithm can make no statement about whether the instance is positive or negative~\cite[Section 3.3]{gms-pew-14}.
We note that, in any of our datasets, we neither found a case of \HTf yielding a false-positive result nor a case of a \HT verification failing.
The asymptotic running time of \HTf is bounded by $O(n^6)$ and the additional verification of \HT adds a further factor of $n$~\cite{gms-pew-14}.

We combine the \cplan datasets that were previously used for evaluations on \HT and \ILP to form the set \dsold~\cite{cgj-cmc-08,ck-sts-12,gms-pew-14}.
We apply the preprocessing rules of Gutwenger at al.~\cite{gms-pew-14} to all instances and discard instances that become trivial, non-planar or cluster-connected, since the latter are easy to solve \cite{cdbf-cpo-08}.
This leaves {\dsoldsize} instances; see \Cref{tab:datasets}.
To create the larger dataset \dsmedncp, we used existing methods from the OGDF to generate instances with up to 500 vertices and up to 50 clusters.
This yields \num{15750} instances, {\dsmedncpsize} out of which are non-trivial after preprocessing.
As this dataset turned out to contain only \qty{10}{\percent} yes-instances, we implemented a new clustered-planar instance generator that is guaranteed to yield yes-instances.
We use it on random planar graphs with up to \num{1000} vertices to generate \num{6300} clustered-planar instances with up to 50 clusters.
Out of these, {\dsmedsize} are non-trivial after preprocessing and make up our dataset~\dsmed.
We provide full details on the generation of our dataset at the end of this section.

We run our experiments on Intel Xeon E5-2690v2 CPUs (3.00 GHz, 10 Cores, 25 MB Cache) with a memory usage limit of 6 GB. As all implementations are single-threaded, we run multiple experiments in parallel using one core per experiment.
This allows us to test more instances while causing a small overhead which affects all implementations in the same way.
The machines run Debian 11 with a 5.10 Linux Kernel.
All binaries are compiled statically using gcc 10.2.1 with flags \texttt{-O3 -march=native} and link-time optimization enabled.
We link against slightly modified versions of OGDF 2022.02 and the PC-tree implementation by Fink et al.~\cite{fpr-eco-21}.
The source code of our implementation and all modifications are available at \href{https://github.com/N-Coder/syncplan}{\texttt{github.com/N-Coder/syncplan}}, \footnote{It is also archived at Software Heritage with ID \href{https://archive.softwareheritage.org/swh:1:snp:0dae4960cc1303cc3575cf04924e19d664f8ad87;origin=https://github.com/N-Coder/syncplan}{\texttt{{swh:1:snp:0dae4960cc1303cc3575cf04924e19d664f8ad87}}}.}
while our dataset is on \href{https://doi.org/10.5281/zenodo.7896021}{Zenodo with DOI \texttt{10.5281/zenodo.7896021}}.

\begin{figure}
  \hfill \begin{subfigure}[t]{0.25\linewidth}
    \centering
    \includegraphics[width=\linewidth,keepaspectratio]{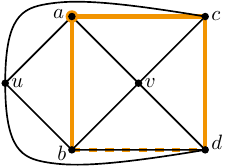}
    \vspace{-1cm}
    \caption{\hspace*{\fill}}
    \label{fig:dfs-subtree-cluster-separated}
  \end{subfigure}\hfill \begin{subfigure}[t]{0.25\linewidth}
    \centering
    \includegraphics[width=\linewidth,keepaspectratio]{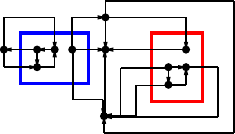}
    \vspace{-1cm}
    \caption{\hspace*{\fill}}
    \label{fig:ht-err-minor}
  \end{subfigure}\hspace*{\fill}\caption{
    \textbf{(a)} Converting the subtree $\{a,b,c,d\}$ with root $a$ (shown in orange) into a cluster will separate vertices $u$ and $v$, as the edge $bd$ (dashed) will also be part of the cluster.
    \textbf{(b)} A clustered-planar graph with two clusters (in addition to the root cluster) that \HT classifies as ``nonCPlanarVerified''.
  }
\end{figure}

\subsubsection*{Details on Dataset Generation}
The dataset \dsold is comprised of the datasets \texttt{P-Small}, \texttt{P-Medium}, \texttt{P-Large} by Chimani and Klein~\cite{ck-sts-12} together with \texttt{PlanarSmallR} (a version of \texttt{Planar}\-\texttt{Small} \cite{cgj-cmc-08} with preprocessing applied), \texttt{PlanarMediumR} and \texttt{PlanarLargeR} by Gutwenger et al.~\cite{gms-pew-14}.
The preprocessing reduced the dataset of Chimani and Klein~\cite{ck-sts-12} to 64 non-trivial instances, leading to dataset \dsold containing {\dsoldsize} instances in total.

The OGDF library can generate an entirely random clustering by selecting random subsets of vertices.
It can also generate a random clustered-planar and cluster-connected clustering on a given graph by running a depth-first search that is stopped at random vertices, forming new clusters out of the discovered trees.
To generate non-cluster-connected but clustered-planar instances, we temporarily add the edges necessary to make a disconnected input graph connected.
For the underlying graphs of \dsmedncp, we use the OGDF to generate three instances for each combination of 
$n\in\{100,200,300,400,500\}$ nodes,
$m\in\{n, 1.5n, 2n, 2.5n, 3n-6\}$ edges, and
$d\in\{10, 20, 30, 40, 50\}$ distinct connected components.
For each input graph, we generate six different clusterings, three entirely random and three random clustered-planar, with
$c\in\{3, 5, 10, 20, 30, 40, 50\}$ clusters.
This yields \num{15750} instances, {\dsmedncpsize} out of which are non-trivial after preprocessing.

It turns out that roughly \qty{90}{\percent} of these instances are not clustered-planar (see \Cref{tab:dsmedncp-results}), even though half of them are generated by a method claiming to only generate clustered-planar instances.
This is because the random DFS-subtree used for clusters by the OGDF only ensures that the generated cluster itself, but not its complement are connected.
Thus, if the subgraph induced by the selected vertices contains a cycle, this cycle may separate the outside of the cluster; see \Cref{fig:dfs-subtree-cluster-separated}.
To reliably generate yes-instances, we implemented a third method for generating random clusterings.
We first add temporary edges to connect and triangulate the given input graph.
Afterwards, we also generate a random subtree and contract it into a cluster.
Each visited vertex is added to the tree with a probability set according to the desired number of vertices per cluster.
To ensure the non-tree vertices remain connected, we only add vertices to the tree whose contraction leaves the graph triangulated, i.e., that have at most two neighbors that are already selected for the tree.
We convert the selected random subtrees into clusters and contract them for the next iterations until all vertices have been added to a cluster.

As we do not need multiple connected components to ensure the instance is not cluster-connected for our \cplan instance generator, we used fewer steps for the corresponding parameter, but extended the number of nodes up to \num{1000} for \dsmed.
The underlying graphs are thus comprised of three instances for each combination of 
$1\leq n\leq1000$ nodes with $0 \equiv n \mod 100$ (i.e.~$n$ is a multiple of 100),
$m\in\{n, 1.5n, 2n, 2.5n, 3n-6\}$ edges, and
$d\in\{1, 10, 25, 50\}$ distinct connected components.
For each input graph, we generate three random clustered-planar clusterings with an expected number of
$c\in\{3, 5, 10, 20, 30, 40, 50\}$ clusters.
This yields 6300 instances which are guaranteed to be clustered-planar, {\dsmedsize} out of which are non-trivial after preprocessing and make up our dataset \dsmed.

\begin{table}
  \centering
  \resizebox*{\textwidth}{!}{
    \sisetup{mode = math, table-format=4}
    \begin{tabular}{r|SSSS|SSSS|SSSS}
          & \multicolumn{4}{c|}{\dsold} & \multicolumn{4}{c|}{\dsmedncp} & \multicolumn{4}{c}{\dsmed} \\
          & \ILP & \HT & \HTf & \SPd & \ILP & \HT  & \HTf & \SPd  & \ILP & \HT  & \HTf & \SPd \\\hline
      Y   & 732  & 792 & 792  & 792  & 181  & 1327 & 1534 & 1535  & 953  & 762  & 2696 & 5170 \\
      N   & 800  & 851 & 851  & 851  & 946  & 6465 & 6463 & 12308 & 0    & 85   & 85   & 0    \\
      ERR & 0    & 0   & 0    & 0    & 5214 & 0    & 0    & 0     & 1263 & 0    & 0    & 0    \\
      TO  & 111  & 0   & 0    & 0    & 7502 & 6051 & 5846 & 0     & 2955 & 4324 & 2390 & 1
    \end{tabular}
  }
\caption{Counts of the results `yes', `no', `error', and `timed out' on \dsold, \dsmedncp and \dsmed.}
  \label{tab:dsmedncp-results}
\end{table}

\begin{figure}
  \begin{subfigure}{.5\linewidth}
    \includegraphics[width=\linewidth]{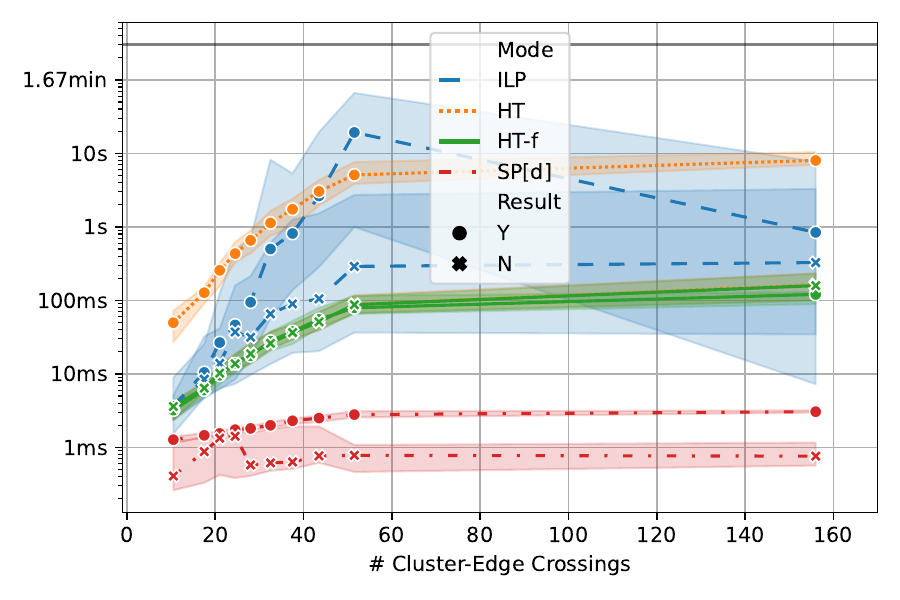}
    \vspace{-1cm}
    \caption{\hspace*{\fill}}
  \end{subfigure}\begin{subfigure}{.5\linewidth}
    \includegraphics[width=\linewidth]{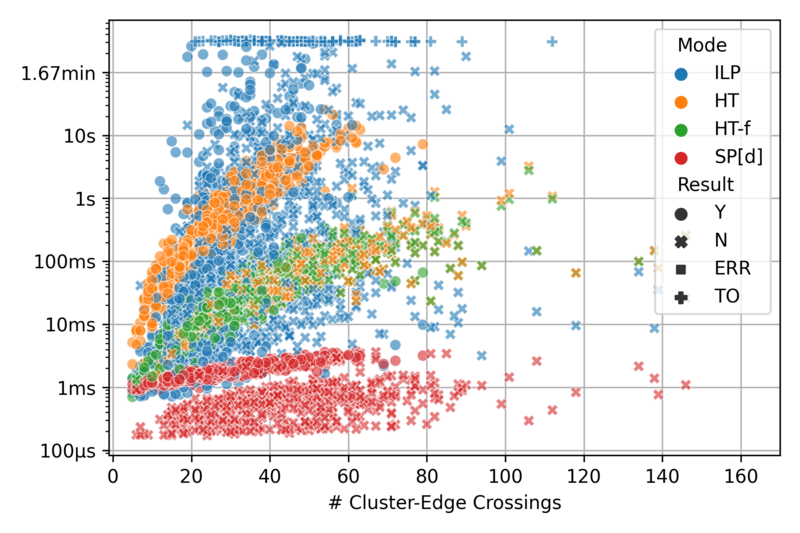}
    \vspace{-1cm}
    \caption{\hspace*{\fill}}
  \end{subfigure}
  \caption{
    Median running times on dataset \dsold \textbf{(a)} together with the underlying scatter plot~\textbf{(b)}.
    For each algorithm, we show running times for yes- and no-instances separately.
    Markers show medians of bins each containing \qty{10}{\percent} of the instances.
    Shaded regions around each line show inter-quartile ranges.
  }
  \label{fig:stats-dsold-time_ns}
\end{figure}

\subsection{Results.}
\Cref{tab:dsmedncp-results} shows the results of running the different algorithms.
The dataset \dsold is split in roughly equal halves between yes- and no-instances and all algorithms yield the same results, except for the~111 instances for which the ILP ran into our 5-minute timeout.
The narrow inter-quartile ranges in \Cref{fig:stats-dsold-time_ns} show that the running time for \HT and \SPd clearly depends on the number of crossings between cluster boundaries and edges in the given instance, while it is much more scattered for \ILP.
Still, all instances with less than 20 such crossings could be solved by \ILP.
For \HT, we can see that the verification and embedding of yes-instances has an overhead of at least an order of magnitude over the non-verifying \HTf.
The running times for \HT on no-instances as well as the times for \HTf on any type of instance are the same, showing that the overhead is solely caused by the verification while the base running time is always the same.
For the larger instances in this test set, \SPd is an order of magnitude faster than \HTf.
For \SPd, we also see a division between yes- and no-instances, where the latter can be solved faster, but also with more scattered running times.
This is probably due to the fact that the test can fail at any (potentially very early) reduction step or when solving the reduced instance.
Furthermore, we additionally generate an embedding for positive instances, which may cause the gap between yes- and no-instances.

\begin{figure}
  \begin{subfigure}{.5\linewidth}
  \includegraphics[width=\linewidth]{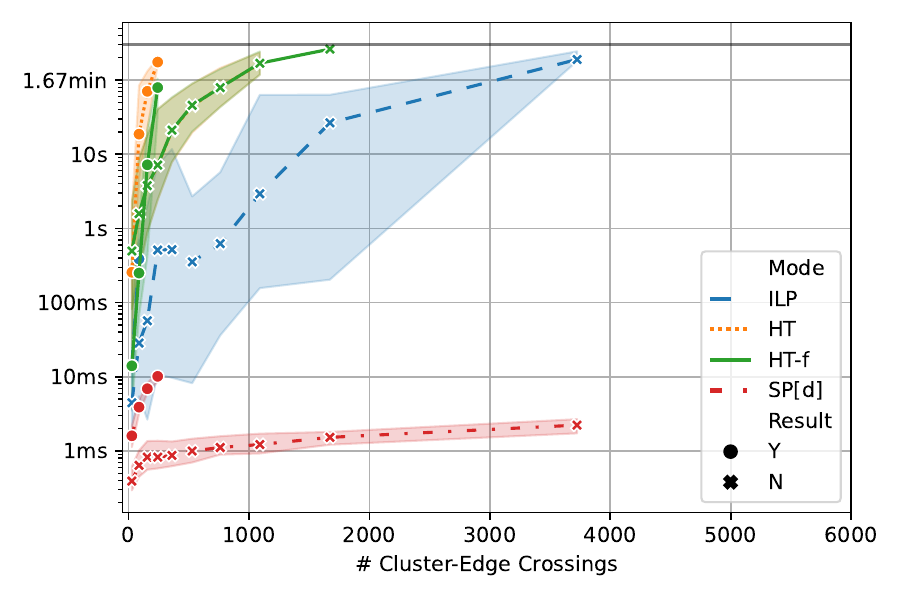}
    \vspace{-1cm}
    \caption{\hspace*{\fill}}
  \end{subfigure}\begin{subfigure}{.5\linewidth}
    \includegraphics[width=\linewidth]{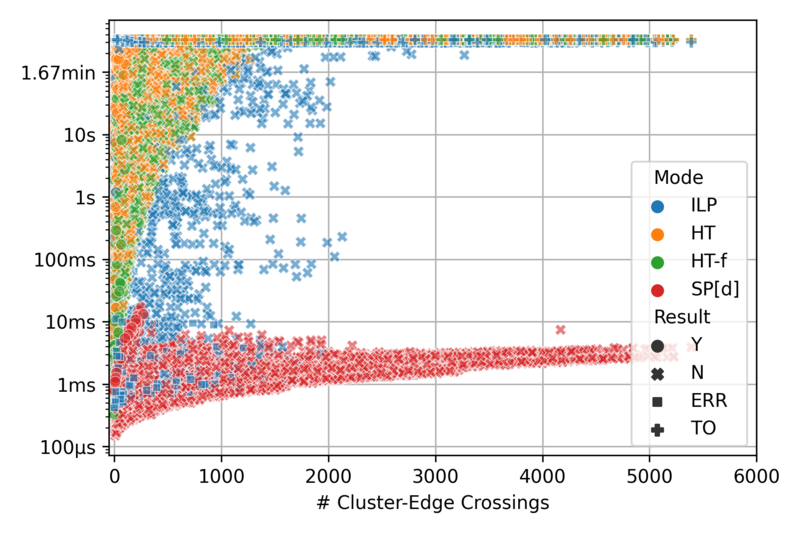}
    \vspace{-1cm}
    \caption{\hspace*{\fill}}
  \end{subfigure}
  \caption{
    Median running times \textbf{(a)} and scatter plot \textbf{(b)} on dataset \dsmedncp.
  }
  \label{fig:stats-clusters-medncp-time_ns}
\end{figure}

The running times on dataset \dsmedncp are shown in \Cref{fig:stats-clusters-medncp-time_ns}.
The result counts in \Cref{tab:dsmedncp-results} show that only a small fraction of the instances are positive.
With only up to \num{300} cluster-edge crossings these instances are also comparatively small.
The growth of the running times is similar to the one already observed for the smaller instances in \Cref{fig:stats-dsold-time_ns}.
\HTf now runs into the timeout for almost all yes-instances of size \num{200} or larger, and both \HT and \HTf time out for all instances of size \num{1500} and larger.
The \ILP only manages to solve very few of the instances, often reporting an ``undefined optimization result for c-planarity computation'' as error; see \Cref{tab:dsmedncp-results}.
The algorithms all agree on the result if they do not run into a timeout or abort with an error, except for one instance that \HT classifies as negative while \SPd found a (positive) solution and also verified its correctness using the generated embedding as certificate.
This is even though the Hanani-Tutte approach by Gutwenger et al.~\cite{gms-pew-14} should answer ``no'' only if the instance truly is negative.
\Cref{fig:ht-err-minor} shows a minimal minor of the instance for which the results still disagree.

\begin{figure}[t]
  \begin{subfigure}{.5\linewidth}
    \includegraphics[width=\linewidth]{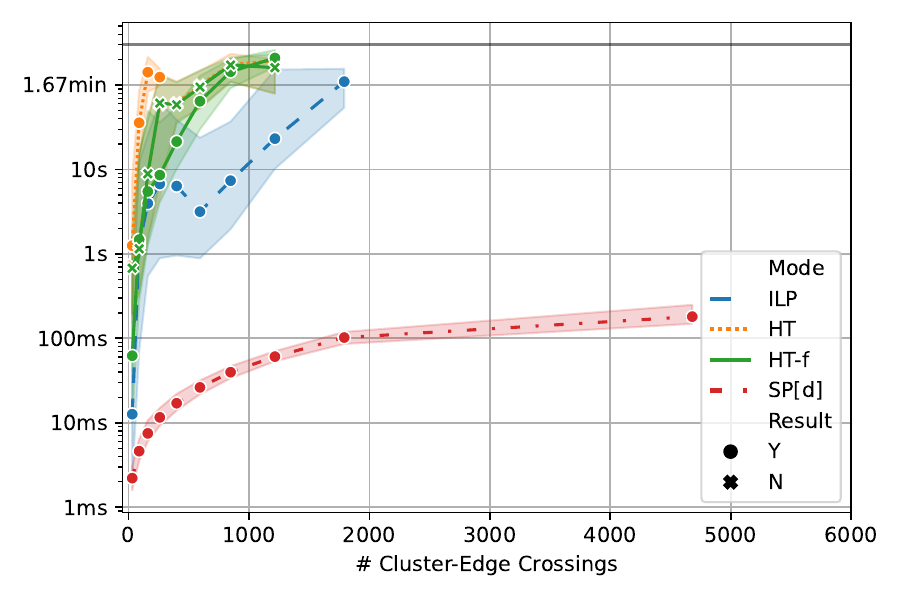}
    \vspace{-1cm}
    \caption{\hspace*{\fill}}
  \end{subfigure}\begin{subfigure}{.5\linewidth}
    \includegraphics[width=\linewidth]{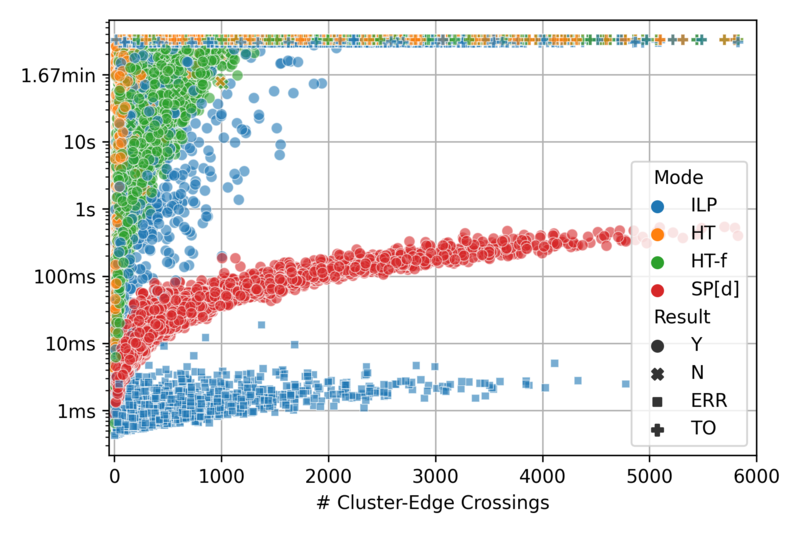}
    \vspace{-1cm}
    \caption{\hspace*{\fill}}
  \end{subfigure}
  \caption{
    Median running times \textbf{(a)} and scatter plot \textbf{(b)} on dataset \dsmed.
  }
  \label{fig:stats-clusters-med-time_ns}
\end{figure}

The running times on dataset \dsmed with only positive instances shown in \Cref{fig:stats-clusters-med-time_ns} are in accordance with the previous results.
We now also see more false-negative answers from the \HT approach, which points to an error in its implementation; see also \Cref{tab:dsmedncp-results}.
The plots clearly show that our approach is much faster than all others.
As the \pqplan reduction fails at an arbitrary step for negative instances,
the running times of positive instances form an upper bound for those of negative instances.
As we also see verifying positive instances to obtain an embedding as far more common use-case, we focus our following engineering on this case.

\section{Engineering Synchronized Planarity}\label{sec:engineering}
In this section, we study how degrees of freedom in the \pqplan algorithm can be used to improve the running times on yes-instances.
The algorithm makes little restriction on the order in which pipes are processed, which gives great freedom to the implementation for choosing the pipe it should process next.
In \Cref{sec:eng-degree,sec:eng-contract-first} we investigate the effects of deliberately choosing the next pipe depending on its degree and whether removing it requires generation of an embedding tree.
As mentioned by the original description of the \pqplan algorithm, there are two further degrees of freedom in the algorithm, both concerning pipes where both endpoints are block-vertices.
The first one is that if both endpoints additionally lie in different connected components, we may apply either \propagate or (\texttt{EncapsulateAnd})\texttt{Join} to remove the pipe.
Joining the pipe directly removes it entirely instead of splitting it into multiple smaller ones, although at the cost of generating larger connected components.
The second one is for which endpoint of the pipe to compute an embedding tree when applying \propagate.
Instead of computing only one embedding tree, we may also compute both at once and then use their intersection.
This preempts multiple following operations propagating back embedding information individually for each newly-created smaller pipe.
We investigate the effect of these two decisions in \Cref{sec:eng-join,sec:eng-intersect}.
Lastly, we investigate an alternative method for computing embedding trees in \Cref{sec:eng-batch}, where we employ a more time-consuming algorithm that in return yields embedding trees for all vertices of a biconnected component simultaneously instead of just for a single vertex.

\begin{figure}[htb]
  \centering
  \includegraphics[width=.8\linewidth]{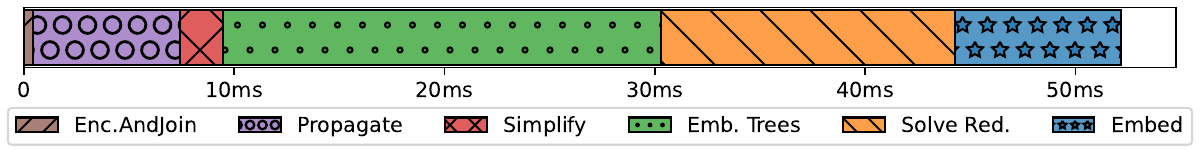}
\caption{
    Average time spent on different operations for \SPd on \dsmed.
  }
  \label{fig:barplot-initial}
\end{figure}

To gain an initial overview over which parts could benefit the most from improvements, \Cref{fig:barplot-initial} shows how the running time is distributed across different operations, averaged over all instances in \dsmed.
It shows that with more than 20ms, that is roughly \qty{40}{\percent} of the overall running time, a large fraction of time is spent on generating embedding trees, while the actual operations contribute only a minor part of roughly \qty{18}{\percent} of the overall running time.
\qty{27}{\percent} of time is spent on solving and embedding the reduced instance and \qty{15}{\percent} is spent on undoing changes to obtain an embedding for the input graph.
Thus, the biggest gains can probably be made by reducing the time spent on generating embedding information in the form of embedding trees.
We use this as rough guideline in our engineering process.

\begin{figure}
  \centering
  \begin{subfigure}{.5\linewidth}
    \includegraphics[width=\linewidth]{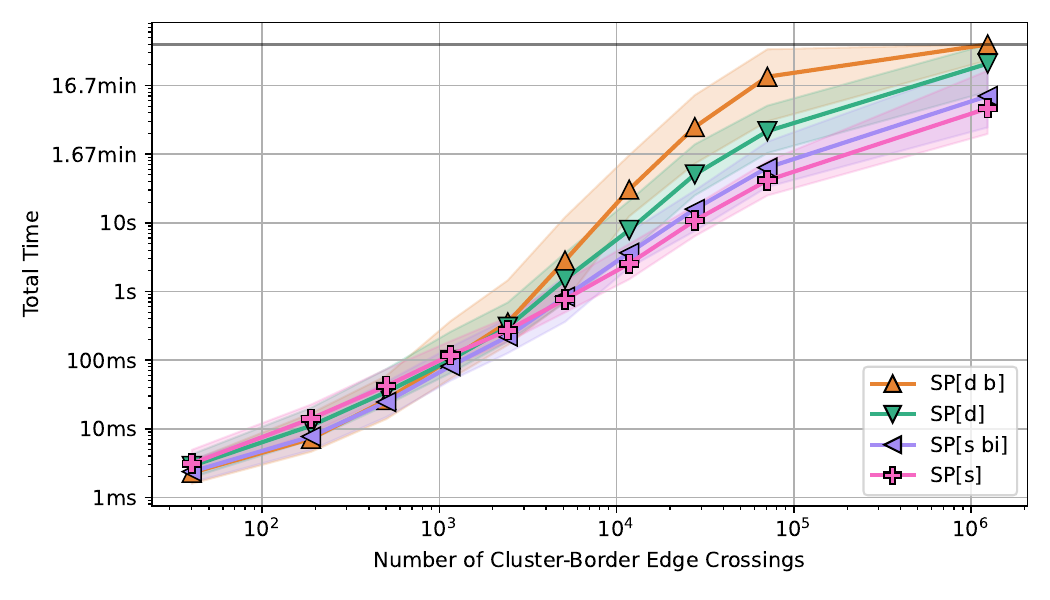}
    \vspace{-1cm}
    \caption{\hspace*{\fill}}
    \label{fig:eng-absolute}
  \end{subfigure}\hfill \begin{subfigure}{.5\linewidth}
    \includegraphics[width=\linewidth]{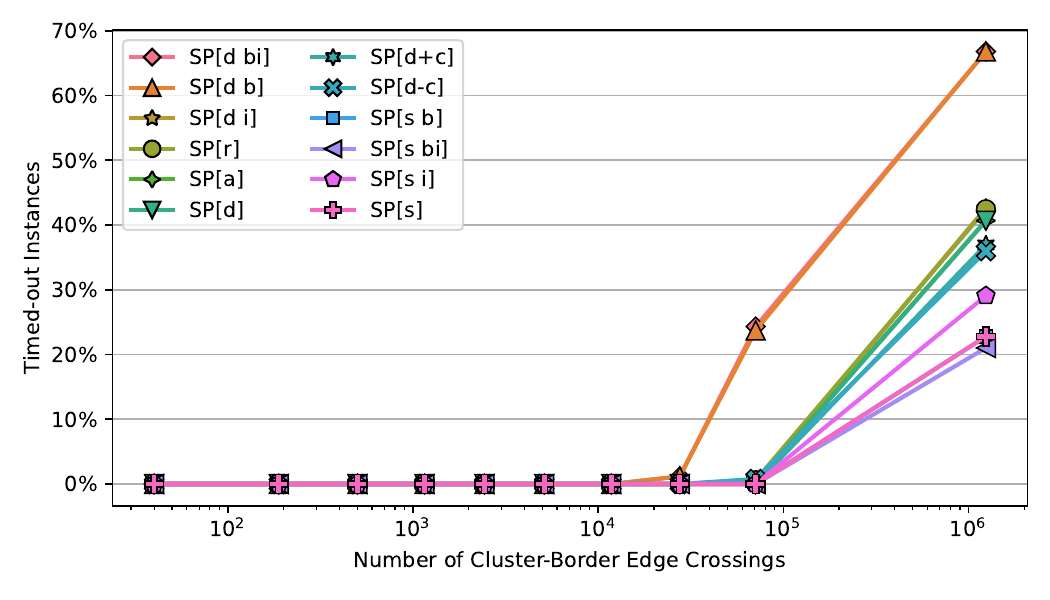}
    \vspace{-1cm}
    \caption{\hspace*{\fill}}
    \label{fig:eng-timeouts}
  \end{subfigure}
  \caption{
    \dslarge median absolute running times \textbf{(a)} and fraction of timeouts \textbf{(b)}.
    Each marker again corresponds to a bin containing \qty{10}{\percent} of the instances.
  }
\end{figure}

\begin{figure}
  \centering
\includegraphics[width=.5\linewidth]{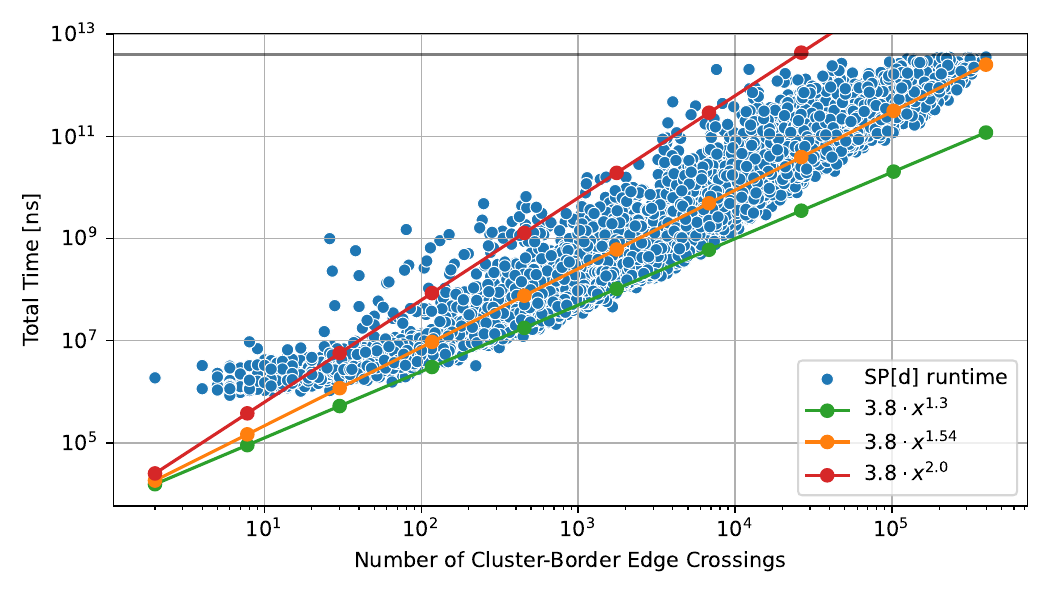}
\caption{
    Scatterplot and estimate for \SPd running time growth behavior on \dslarge.
  }
  \label{fig:poly-fit}
\end{figure}

\subsubsection*{Dataset Generation}
To tune the running time of our algorithm on larger instances, we increased the size of the generated instances by a factor of 100 by changing the parameters of our own cluster-planar instance generator to
$n\in\{$\num{100}, \num{500}, \num{1000}, \num{5000}, \num{10000}, \num{50000}, \num{100000}$\},$
$d\in\{1, 10, 100\},$
$c\in\{3, 5, 10, 25, 50, 100, \num{1000}\}$ for dataset \dslarge.
This yields 6615 instances, out of which {\dslargesize} are non-trivial after preprocessing; see~\Cref{tab:datasets}.

In addition to the \cplan dataset we also generate a dataset that uses the reduction from \consefe.
We do so by generating a random connected and planar embedded graph as shared graph.
Each exclusive graph contains further edges which are obtained by randomly splitting the faces of the embedded shared graph until we reach a desired density.
For the shared graphs, we generate three instances for each combination of
$n\in\{$\num{100}, \num{500}, \num{1000}, \num{2500}, \num{5000}, \num{7500}, \num{10000}$\}$ nodes and
$m\in\{n, 1.5n, 2n, 2.5n\}$ edges.
For $d\in\{0.25, 0.5, 0.75, 1\}$,
we then add $(3n-6-m)\cdot d$ edges to each exclusive graph, i.e., the fraction $d$ of the number of edges that can be added until the graph is maximal planar.
We also repeat this process three times with different initial random states for each pair of shared graph and parameter $d$.
This leads to the dataset \dssefe containing {\dssefesize} instances.

We also generate a dataset of \pqplan instances by taking a random planar embedded graph and adding pipes between vertices of the same degree, using a bijection that matches their current rotation.
The underlying graphs are comprised of three instances for each combination of 
$n\in\{$\num{100}, \num{500}, \num{1000}, \num{5000}, \num{10000}, \num{50000}, \num{100000}$\}$ nodes,
$m\in\{1.5n, 2n, 2.5n\}$ edges, and
$d\in\{1, 10, 100\}$ distinct connected components.
Note that we do not include graphs that would have no edges, e.g., those with $n=100$ and $d=100$.
For each input graph, we generate three random \pqplan instances with 
$p\in\{0.05n, 0.1n, 0.2n\}$ pipes.
This leads to the dataset \dspq containing {\dspqsize} instances.

Altogether, our six datasets contain {\dstotalsize} instances in total.
For the test runs on these large instances, we increase the timeout to 1~hour.

\Cref{fig:eng-absolute} shows the result of running our baseline variant \SPd of the \pqplan algorithm (together with selected further variants of the algorithm from subsequent sections) on dataset \dslarge.
Note that, because the dataset spans a wide range of instance sizes and thus the running times also span a range of different magnitudes, the plot uses a log scale for both axes.
\Cref{fig:eng-timeouts} shows the fraction of runs that timed out for each variant.
To estimate the practical runtime growth behavior, we also fit a polynomial to the data shown in \Cref{fig:poly-fit} and thereby find the running time growth behavior to be similar to $d^{1.5}$, where $d$ is the number of crossings between edges and cluster borders.

\begin{figure}
  \centering
  \begin{subfigure}{.5\linewidth}
    \includegraphics[width=\linewidth]{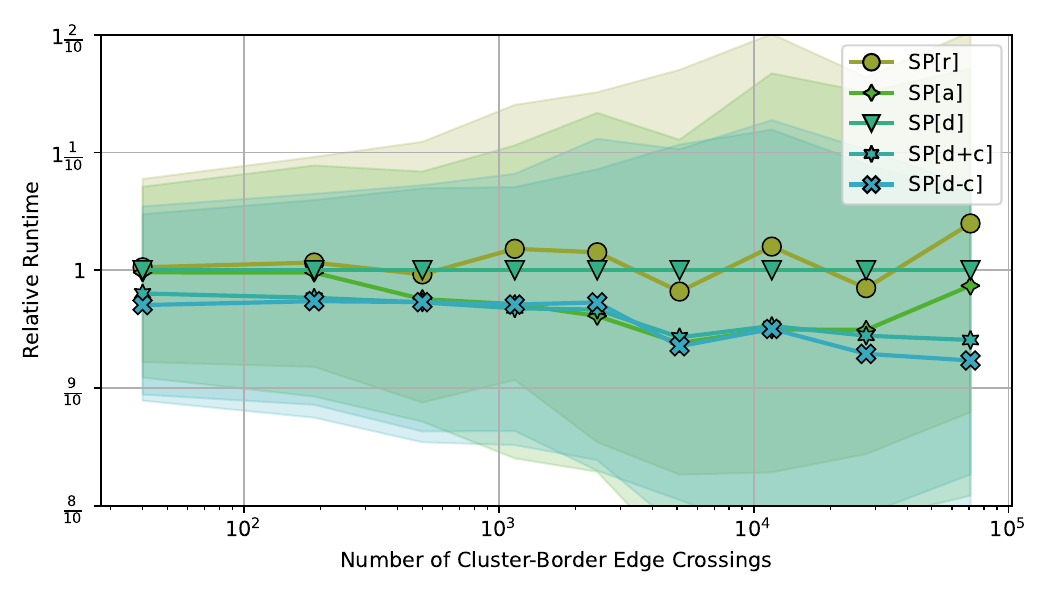}
    \vspace{-1cm}
    \caption{\hspace*{\fill}}
    \label{fig:eng-ordering}
  \end{subfigure}\hfill \begin{subfigure}{.5\linewidth}
    \includegraphics[width=\linewidth]{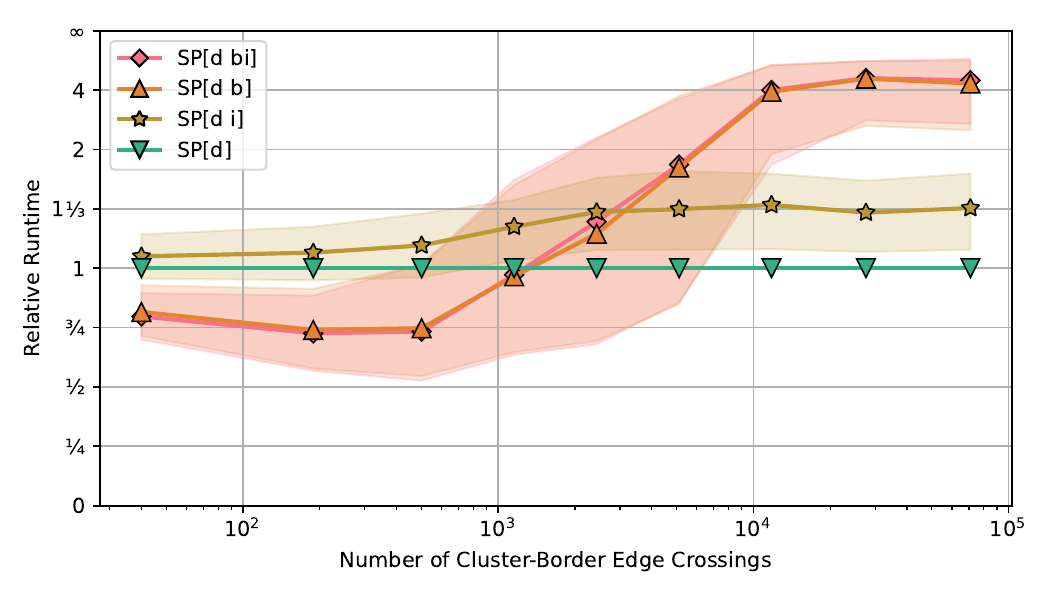}
    \vspace{-1cm}
    \caption{\hspace*{\fill}}
    \label{fig:eng-bi}
  \end{subfigure}
  \caption{
    Relative running times when \textbf{(a)}~sorting by pipe degree or applicable operation and
    \textbf{(b)}~when handling pipes between block-vertices via intersection or join.
    Note the different scales on the y-axis.
  }
\end{figure}

\subsection{Pipe Ordering}\label{sec:eng-degree}\label{sec:eng-contract-first}\label{sec:eng-order}
To be able to deliberately choose the next pipe, we keep a heap of all pipes in the current instance, where the ordering function can be configured.
Note that the topmost pipe from this heap may not be feasible, in which case we will give priority to the close-by pipe of higher degree that blocks the current pipe from being feasible (see \cite[Lemma 3.5]{bfr-spw-21}).
We compare the baseline variant \SPd sorting by descending (i.e. largest first) degree with the variant~\combconf{a} sorting by ascending degree, and~\combconf{r} using a random order.
Note that for these variants, the ordering does not depend on which operation is applicable to a pipe or whether this operation requires the generation of an embedding tree.
To see whether making this distinction affects the running time, we also compare the variants~\combconf{d+c}, which prefers to process pipes on which \contract can be applied, and~\combconf{d-c}, which defers such pipes to the very end, processing pipes requiring the generation of embedding trees first.

To make the variants easier to compare, \Cref{fig:eng-ordering} shows running times relative to that of the baseline \combconf{d}.
Note that we do not show the median of the last bin, in which up to \qty{70}{\percent} of the runs timed out, while this number is far lower for all previous bins; see \Cref{fig:eng-timeouts}.
\Cref{fig:eng-ordering} shows that the median running times differ by less than \qty{10}{\percent} between these variants.
The running time of~\combconf{r} seems to randomly alternate between being slightly slower and slightly faster than \SPd.
\SPd is slightly slower than \combconf{a} for all bins except the very first and very last, indicating a slight advantage of processing small pipes before bigger ones on these instances.
Interestingly, \SPd is also slower than both \combconf{d+c} and \combconf{d-c} for all bins.
The fact that these two variants have the same speed-ups indicates that \contract should not be interleaved with the other operations, while it does not matter whether it is handled first or last.
Still, the variance in relative running times is high and none of the variants is consistently faster on a larger part of the instances.
To summarize, the plots show a slight advantage for not interleaving operation \contract with the others or sorting by ascending degree, but this advantage is not significant in the statistical sense; see \Cref{sec:stat-sign}.
We keep \SPd as the baseline for our further analysis.

\subsection{Pipes with two Block-Vertex Endpoints}\label{sec:eng-join}\label{sec:eng-intersect}
Our baseline always processes pipes where both endpoints are block-vertices by applying \propagate or \simplify based on the embedding tree of an arbitrary endpoint of the pipe.
Alternatively, if the endpoints lie in different connected components, such pipes can also be joined directly by identifying their incident edges as in the second step of \contract.
This directly removes the pipe entirely instead of splitting into further smaller pipes, although it also results in larger connected components.
We enable this joining in variant \combconf{d b}.
As a second alternative, we may also compute the embedding trees of both block-vertices and then propagate their intersection.
This preempts the multiple following operations propagating back embedding information individually for each newly-created smaller pipe.
We enable this intersection in variant \combconf{d i}.
Variant~\combconf{d bi} combines both variants, preferring the join and only intersecting if the endpoints are in the same connected component.
We compare the effect of differently handling pipes with two block-vertex endpoints in variants~\combconf{d b},~\combconf{d i} and~\combconf{d bi} with the baseline \SPd, which computes the embedding tree for an arbitrary endpoint and only joins pipes where both pipes are cut-vertices.

\Cref{fig:eng-bi} shows that \combconf{d b} (and similarly \combconf{d bi}) is faster by close to \qty{25}{\percent} on instances with less than \num{1000} cluster-border edge crossings, but quickly grows 5 times slower than \SPd for larger instances.
This effect is also visible in the absolute values of \Cref{fig:eng-absolute}.
This is probably caused by the larger connected components (see the last column of \Cref{tab:opstats}), which make the computation of embedding trees more expensive.
Only inserting an embedding tree instead of the whole connected component makes the embedding information of the component directly available in a compressed form without the need to later process the component in its entirety again.
\Cref{fig:eng-bi} also shows that~\combconf{d i} is up to a third slower than \SPd, indicating that computing both embedding trees poses a significant overhead while not yielding sufficiently more information to make progress faster.
We also evaluated combinations of the variants from this section with the different orderings from the previous section, but observed no notable differences in running time behavior.
The effects of the variants from this section always greatly outweigh the effects from the different orderings.
To summarize, as the plots only show an advantage of differently handling pipes between block-vertices for small instances, but some strong disadvantages especially for larger instances, we keep \SPd as our baseline.

\begin{figure}
  \centering
  \begin{subfigure}{.5\linewidth}
    \includegraphics[width=\linewidth]{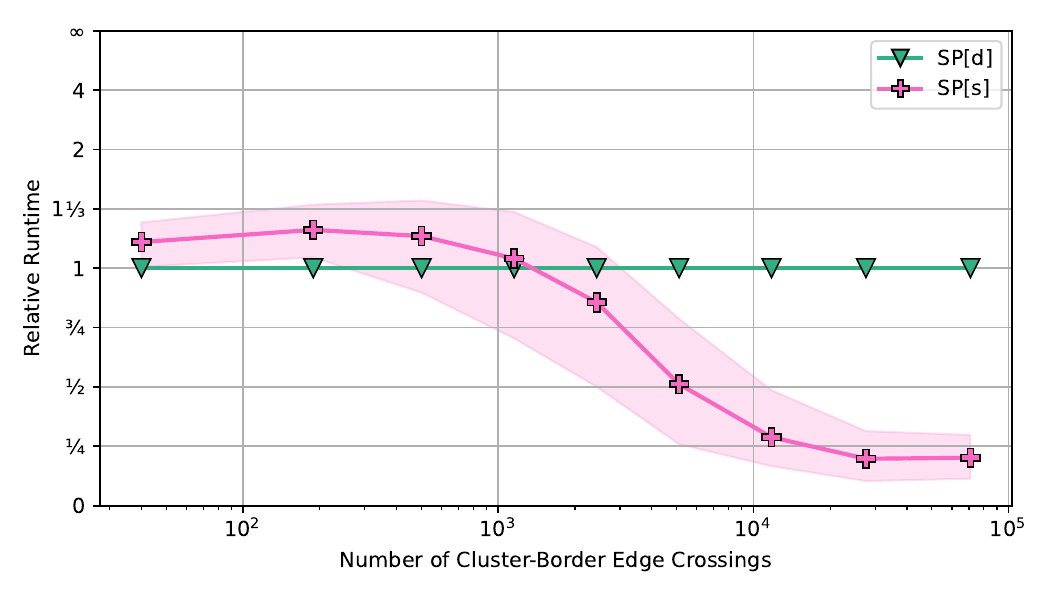}
    \vspace{-1cm}
    \caption{\hspace*{\fill}}
    \label{fig:eng-spqr}
  \end{subfigure}\hfill \begin{subfigure}{.5\linewidth}
    \includegraphics[width=\linewidth]{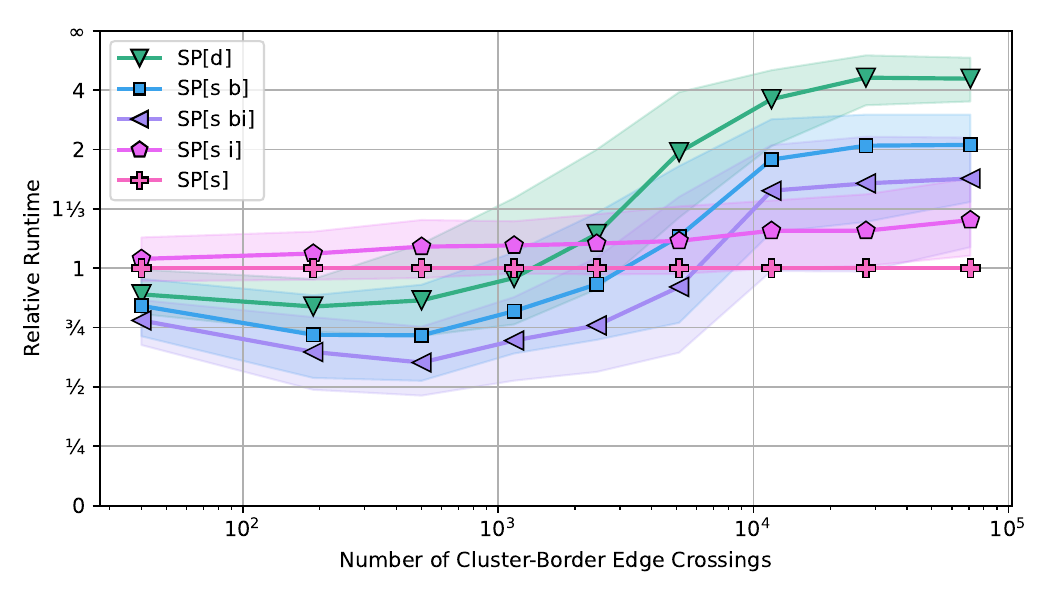}
    \vspace{-1cm}
    \caption{\hspace*{\fill}}
    \label{fig:eng-spqr-var}
  \end{subfigure}
  \caption{
    Relative running times for \textbf{(a)} SPQR-tree batched embedding tree generation and
    \textbf{(b)} for different variants thereof.
  }
\end{figure}

\subsection{Batched Embedding Tree Generation}\label{sec:eng-batch}
Our preliminary analysis showed that the computation of embedding trees consumes a large fraction of the running time (see \Cref{fig:barplot-initial}),
which cannot be reduced significantly by using the degrees of freedom of the algorithm studied in the previous two sections.
To remedy the overhead of recomputing embedding trees multiple times we now change the algorithm to no longer process pipes one-by-one, but to process all pipes of a biconnected component in one batch.
This is facilitated by an alternative approach for generating embedding trees not only for a single vertex, but for all vertices of a biconnected component.
The embedding tree of a vertex $v$ can be derived from the SPQR-tree using the approach described by Bläsius et al.~\cite{br-spo-16}:
Each occurrence of $v$ in a ``parallel'' skeleton of the SPQR-tree corresponds to a (PQ-tree) P-node in the embedding tree of $v$, each occurrence in a ``rigid'' to a (PQ-tree) Q-node.
This derivation can be done comparatively quickly, in time linear in the degree of~$v$.
Thus, once we have the SPQR-tree of a biconnected component available, we can apply all currently feasible \propagate and \simplify operations in a single batch with little overhead.
The SPQR-tree computation takes time linear in the size of the biconnected component, albeit with a larger linear factor than for the linear-time planarity test that yields only a single embedding tree.
In a direct comparison with the planarity test, this makes the SPQR-tree the more time-consuming approach.

We enable the batched embedding tree computation based on SPQR-trees in variant \combconf{s}.
\Cref{fig:eng-spqr,fig:eng-absolute} show that for small instances, this yields a slowdown of close to a third.
Showing a behavior inverse to~\combconf{d b}, \combconf{s} grows faster for larger instances and its speed-up even increases to up to 4 times as fast as the baseline \SPd.
This makes~\combconf{s} the clear champion of all variants considered so far.
We will thus use it as baseline for our further evaluation, where we combine \combconf{s} with other, previously considered flags.

\subsection{SPQR-Batch Variations}\label{sec:eng-combinations}
\Cref{fig:eng-spqr-var} switches the baseline between the two variants shown in \Cref{fig:eng-spqr} and additionally contains combinations of the variants from \Cref{sec:eng-join} with the SPQR-batch computation.
As in \Cref{fig:eng-bi}, the intersection of embedding trees in \combconf{s i} is consistently slower, albeit with a slightly smaller margin.
The joining of blocks in \combconf{s b} also shows a similar behavior as before, starting out \qty{25}{\percent} faster for small instances and growing up to \qty{100}{\percent} slower for larger instances.
Again, this is probably because too large connected components negatively affect the computation of SPQR-trees.
Still, the median of \combconf{s b} is consistently faster than \SPd.
Different to before, \combconf{s bi} is now faster than \combconf{s b}, making it the best variant for instances with up to \num{5000} cluster-border edge crossings.
This is probably because in the batched mode, there is no relevant overhead for obtaining a second embedding tree, while the intersection does preempt some following operations.
To summarize, for instances up to size \num{5000}, \combconf{s bi} is the fastest variant, which is outperformed by \combconf{s} on larger instances.
This can also be seen in the absolute running times in \Cref{fig:eng-absolute}, where \combconf{s} is more than an order of magnitude faster than \combconf{d b} on large instances.

\section{Further Analysis}\label{sec:further-analysis}
\begin{figure}
  \begin{subfigure}{\linewidth}
    \centering
    \includegraphics[width=.95\linewidth]{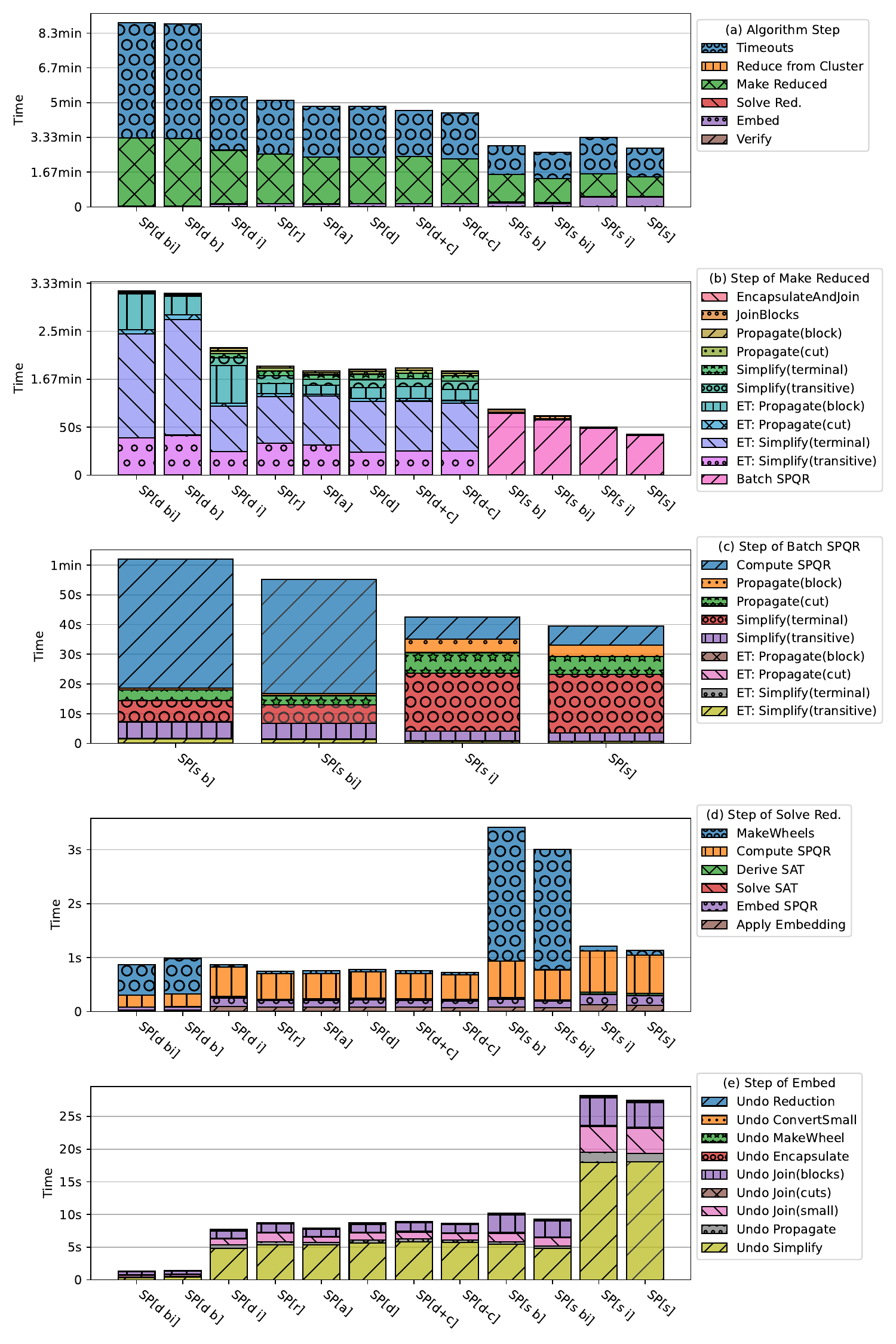}
    \phantomsubcaption\label{fig:barplots-all}
    \phantomsubcaption\label{fig:barplots-makeReduced}
    \phantomsubcaption\label{fig:barplots-batchSPQR}
    \phantomsubcaption\label{fig:barplots-solveReduced}
    \phantomsubcaption\label{fig:barplots-embed}
  \end{subfigure}
  \vspace{-1cm}
  \caption{The average running time of our different \pqplan variants.}
  \label{fig:barplots}
\end{figure}

In this section, we provide further in-depth analysis of the different variants from the previous section and also analyze their performance on the remaining datasets to give a conclusive judgement.
To gain more insights into the runtime behavior, we measured the time each individual step of the algorithm takes when using the different variants.
An in-depth analysis of this data is given in \Cref{apx:eng-opstats}, where \Cref{fig:barplots} also gives a more detailed visualization of per-step timings.
The per-step data corroborates that the main improvement of faster variants is greatly reducing the time spent on the generation of embedding trees, at the cost of slightly increased time spent on the solve and embed phases.

To further verify our ranking of variants' running times from the previous sections, we also use a statistical test to check whether one variant is significantly faster than another.
The results presented in \Cref{sec:stat-sign} corroborate our previous results, showing that pipe ordering has no significant effect while the too large connected components and batched processing of pipes using SPQR-trees significantly change the running time.

The results of the remaining datasets \dssefe and \dspq are presented in \Cref{asec:other-ds} and mostly agree with the results on \dslarge, with \combconf{d b} clearly being the slowest and \combconf{s} being the fastest on large instances.
The main difference is the magnitude of the overhead generated by large connected components for variants with flag \combflag{b}.

\newcommand{\R}[1]{\multicolumn{1}{l}{\rlap{\rotatebox{50}{\parbox{2cm}{\raggedright \setstretch{0.7}\hangindent=1.7ex\hangafter=1#1}}}}}
\begin{table}[tbp]
  \centering
  \resizebox*{\linewidth}{!}{
    \begin{tabular}{l|S|SSS|SSSS|SS[table-format=4]|S[table-format=5]}
      \R{Mode} & \R{Total Time} &
      \R{Make Reduced} & \R{Solve Reduced} & \R{Embed} & 
      \R{Enc.And-Join} & \R{Propagate} & \R{Simplify} & \R{Compute Emb.\ Tree} & 
      \R{Undo Simplify} & \R{\#Simplify Operations} & \R{Max.\ Bicon.\ Size} \\ \hline

      \combconf{d}    & 142.68 & 133.08 & 0.82 & 8.78  & 0.25 & 5.00  & 13.79 & 91.34  & 5.64  & 1811 & 2780  \\
      \combconf{d b}  & 197.17 & 194.72 & 0.99 & 1.46  & 0.63 & 1.36  & 1.53  & 186.18 & 0.42  & 652  & 13021 \\
      \combconf{s}    & 86.57  & 57.75  & 1.25 & 27.56 & 0.57 & 9.84  & 22.38 & 7.61   & 18.03 & 2696 & 2890  \\
      \combconf{s b}  & 93.07  & 79.25  & 3.55 & 10.26 & 2.92 & 4.29  & 12.74 & 46.31  & 5.46  & 1421 & 22965 \\
      \combconf{s bi} & 81.32  & 68.90  & 3.09 & 9.32  & 2.51 & 3.79  & 11.52 & 41.16  & 4.84  & 1448 & 23284 \\
      \hline
    \end{tabular}
  }
  \caption{
    Average values for different variants of \SP on dataset \dslarge.
    All values, except for the counts in the last two columns, are running times in seconds.
    The first data column shows the average total running time, followed by how this is split across the three phases.
    The following four columns show the composition of the running time of the ``Make Reduced'' step.
    The last three columns detail information about the ``Undo Simplify'' step in the ``Embed'' phase, and the maximum size of biconnected components in the reduced instance.
  }
  \label{tab:opstats}
\end{table}

\subsection{Detailed Runtime Profiling}\label{apx:eng-opstats}
\Cref{tab:opstats} shows the per-step running time information aggregated for variants studied in the previous section.
\Cref{fig:barplots} in greater detail shows how the running time spent is split on average across the different steps of the algorithm (\Cref{fig:barplots-all}) and then also further drills down on the composition of the individual steps that make the instance reduced (\Cref{fig:barplots-makeReduced}), solve the reduced instance (\Cref{fig:barplots-solveReduced}), and then derive a solution and an embedding for the input instance by undoing all changes while maintaining the embedding (\Cref{fig:barplots-embed}).
For variants that use the SPQR-tree for embedding information generation, we also analyze the time spent on the steps of this batch operation (\Cref{fig:barplots-batchSPQR}).
Note that we do not have these measurements available for runs that timed out.
To ensure that the bar heights still correspond to the actual overall running times in the topmost plot, we add a bar corresponding to the time consumed by timed-out runs on top.
This way, ordering the bars by height yields roughly the same order of variants as we already observed in \Cref{fig:eng-absolute}.

\Cref{fig:barplots-makeReduced} clearly shows that the majority of time during the reduce step is spent on generating embedding information, either in the form of directly computing embedding trees (bars prefixed with ``ET'') or by computing SPQR trees.
This can also be seen by comparing column ``Make Reduced'' in \Cref{tab:opstats} with column ``Compute Emb Tree''.
Only for the fastest variants, those with flag~\combflag{s} and without~\combflag{b}, the execution of the actual operations of the algorithm becomes more prominent over the generation of embedding information in \Cref{fig:barplots-batchSPQR}.
Here, the terminal case of the \simplify operation (described in the bottom left part of \Cref{fig:syncplan-ops}) now takes the biggest fraction of time, and actually also a bigger absolute amount of time than for the other, slower variants with flag~\combflag{b} enabled.
This is probably because, instead of being joined as with flag~\combflag{b} enabled, here pipes between block-vertices are split by \propagate into multiple smaller pipes, which then need to be removed by \simplify.
This leads to the variants without~\combflag{b} needing, on average, roughly two to three times as many \simplify applications as those with~\combflag{b}; see \Cref{tab:opstats}.

The larger biconnected components caused by~\combflag{b} may also be the reason why the insertion of wheels takes a larger amount of time for variants with~\combflag{b} in the solving phase shown in \Cref{fig:barplots-solveReduced}.
When replacing a cut-vertex by a wheel, all incident biconnected components with at least two edges incident to the cut-vertex get merged.
Updating the information stored with the vertices of the biconnected components is probably consuming the most time here, as undoing the changes by contracting the wheels is again very fast. Other than the ``MakeWheels'' part, most time during the solving phase is spent on computing SPQR trees, although both is negligible in comparison to the overall running time.

The running times of the embedding phase given in \Cref{fig:barplots-embed} show an interesting behavior as they increase when the ``Make Reduced'' phase running time decreases, indicating a potential trade-off to be made; see also the ``Embed'' column in  \Cref{tab:opstats}.
As the maximum time spent on the ``Make Reduced'' phase is still slightly larger, variants where this phase is faster while the embedding phase is slower are still overall the fastest.
The biggest contribution of running time in the latter phase is the undoing of \simplify operations, which means copying the embedding of one endpoint of a removed pipe to the other.
The time spent here roughly correlates with the time spent on applying the \simplify operations in the first place (see \Cref{tab:opstats}).

To summarize, the per-step data corroborates that the main improvement of faster variants is greatly reducing the time spent on the generation of embedding trees, at the cost of slightly increased time spent on the solve and embed phases.
Flags~\combflag{s} and~\combflag{b} have the biggest impact on running times, while flag~\combflag{i} and the processing order of pipes do not seem to have a significant influence on the overall running time.
While the variants with~\combflag{s} clearly have the fastest overall running times, there is some trade-off between the amounts of time spent on different phases of the algorithm when toggling the flag~\combflag{b}.

\subsection{Statistical Significance}\label{sec:stat-sign}
To test whether one variant is (in the statistical sense) significantly faster than another, we use the methodology proposed by Radermacher~\cite[Section 3.2]{rad-ggd-20} for comparing the performance of graph algorithms.
For a given graph $G$ and two variants of the algorithm described by their respective running times $f_A(G), f_B(G)$ on $G$,
we want to know whether we have a likelihood at least $p$ that the one variant is faster than the other by at least a factor $\Delta$.
To do so, we use the binomial sign test with advantages as used by Radermacher~\cite{rad-ggd-20}, where we fix two values $p \in [0, 1]$ and $\Delta \geq 1$, and study the following hypothesis given a random graph $G$ from our dataset:
Inequality $f_A(G) \cdot \Delta < f_B(G)$ holds with probability $\pi$, which is at least $p$.
The respective null hypothesis is that the inequality holds with probability less than $p$.
Note that this is an experiment with exactly two outcomes (the inequality holding or not), which we can independently repeat on a sequence of $n$ graphs and obtain the number of instances $k$ for which the inequality holds.
Using the binomial test, we can check the likelihood of obtaining at most $k$ successes by drawing $n$ times from a binomial distribution with probability $p$.
If this likelihood is below a given significance level $\alpha\in[0,1]$, that is the obtained result is unlikely under the null hypothesis, we can reject the null hypothesis that the inequality only holds with a probability less than $p$.

Fixing the significance level to the commonly-used value $\alpha=0.05$, we still need to fix values for $p$ and $\Delta$ to apply this methodology in practice.
We will use three different values for $p\in[0.25, 0.5, 0.75]$, corresponding to the advantage on a quarter, half, and three quarters of the dataset.
To obtain values for $\Delta$, we will split our datasets evenly into two halves $\mathcal G_\text{train}$ and $\mathcal G_\text{verify}$, using the $\mathcal G_\text{train}$ to obtain an estimate for $\Delta$ and $\mathcal G_\text{verify}$ to verify this value.
For a given value of $p$, we set~$\Delta'$ to the largest value such that $f_A(G) \cdot \Delta' < f_B(G)$ holds for $p\cdot|\mathcal G_\text{train}|$ instances.
To increase the likelihood that we can reject the null hypothesis in the verification step on $\mathcal G_\text{verify}$, we will slightly discount the obtained value of $\Delta'$, using $\Delta=\min(1,c\cdot\Delta')$ instead with $c$ set to $0.75$.

\begin{figure}[htb]
  \centering
  \includegraphics[width=\linewidth]{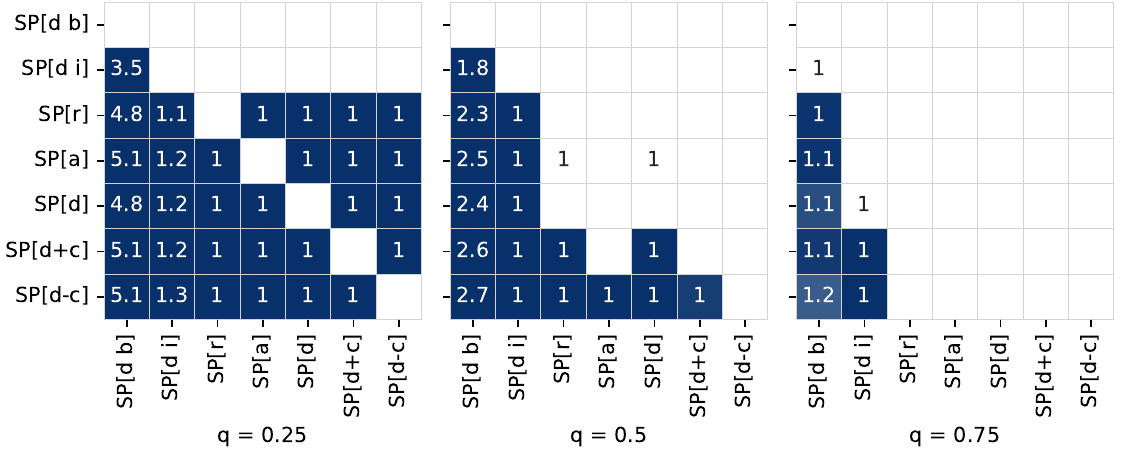}
\caption{
    Advantages of variants without flag \combflag{s} on \dslarge instances of size at least \num{5000}.
    Blue cell backgrounds indicate significant values, while in cells with white background, we were not able to reject the null-hypothesis with significance $\alpha=0.05$.
    Empty cells indicate that the fraction where the one algorithm is better than the other is smaller than $p$.
  }
  \label{fig:eng-binom-bml}
\end{figure}

Applying this methodology, \Cref{fig:eng-binom-bml} compares the pairwise advantages of the variants from \Cref{sec:eng-order,sec:eng-join}.
We see that \combconf{d i} and especially \combconf{d b} are significantly slower than the other variants:
for the quarter of the dataset with the most extreme differences, the advantage rises up to a 5-fold speed-up for other variants,
while slight advantages still persist when considering three quarters of instances.
Conversely, not even on a quarter of instances are \combconf{d i} and \combconf{d b} faster than other variants.
Comparing the remaining variants with each other, we see that each variant has at least a quarter of instances where it is slightly faster than the other variants, but always with no noticeable advantage, that is $\Delta=1$.
This is not surprising as the relative running times are scattered evenly above and below the baseline in \Cref{fig:eng-ordering}.
For half of the dataset, \combconf{d-c} is still slightly faster than other variants,
while no variant from \Cref{sec:eng-order} is faster than another for at least three quarters of instances.
To summarize, our results here corroborate the findings from \Cref{sec:eng-order,sec:eng-join}, with \combconf{d i} and \combconf{d b} as the clearly slowest variants.
While there is no clear winner among the other variants, at least \combconf{d-c} is slightly faster than the others on half of the dataset, but still has no noticeable advantage.

\begin{figure}
  \centering
  \begin{subfigure}{\linewidth}
    \includegraphics[width=\linewidth]{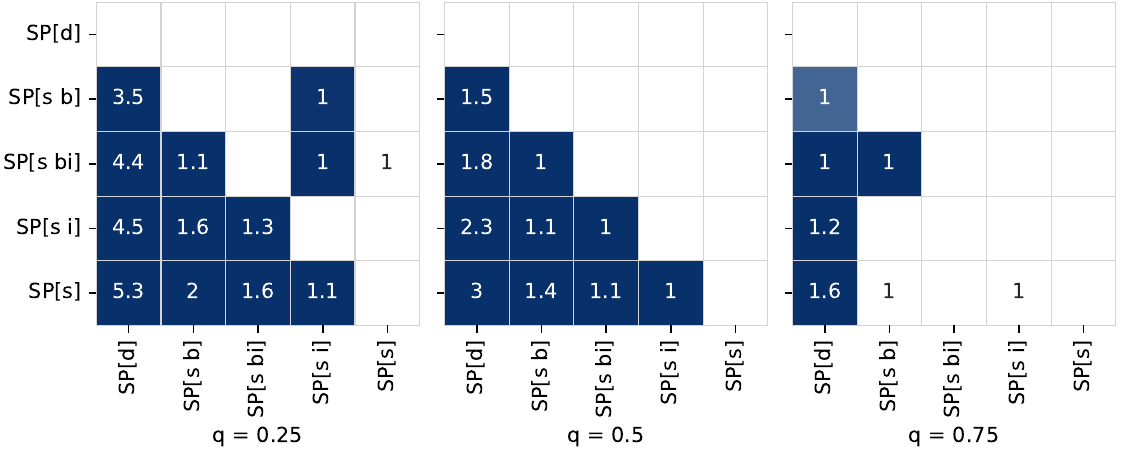}
    \vspace{-1cm}
    \caption{\hspace*{\fill}}
    \label{fig:eng-binom-sml}
  \end{subfigure}\\[.5cm]
  \begin{subfigure}{\linewidth}
    \includegraphics[width=\linewidth]{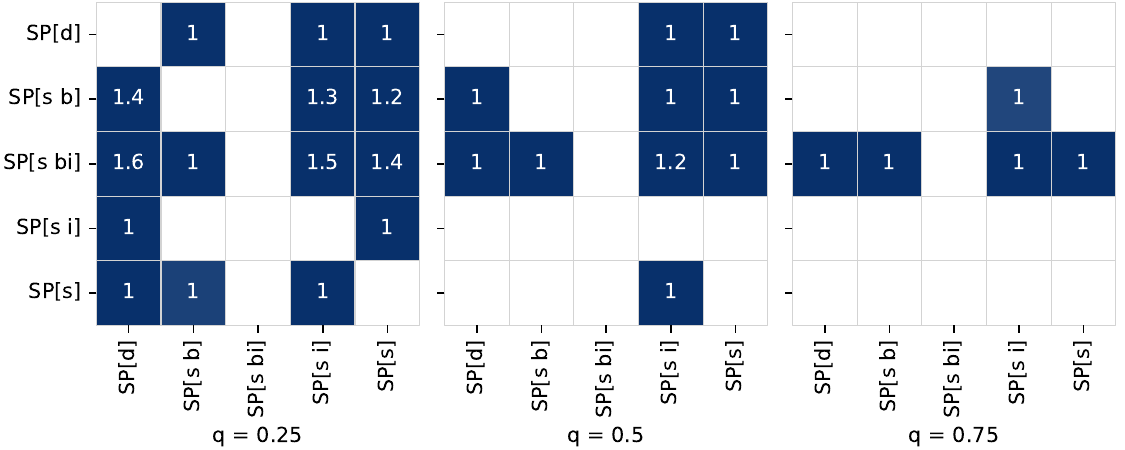}
    \vspace{-1cm}
    \caption{\hspace*{\fill}}
    \label{fig:eng-binom-ss}
  \end{subfigure}
  \caption{
    Advantages of variants with flag \combflag{s} on \dslarge instances of size at least \num{5000} (a) and at most (b) \num{5000}.
  }
\end{figure}

\Cref{fig:eng-binom-sml,fig:eng-binom-ss} compare the pairwise advantages of the variants from \Cref{sec:eng-batch,sec:eng-combinations} (see also \Cref{fig:eng-spqr-var}) for instances with more and less than \num{5000} cluster-border edge crossings, respectively.
For the larger instances of \Cref{fig:eng-binom-sml}, the variants with flag \combflag{s} outperform \SPd on at least \qty{75}{\percent} of instances, with advantages as high as a factor of~5 on at least a quarter of instances.
Furthermore, \combconf{s} outperforms the variants with additional flags \combflag{b} and \combflag{i} on at least half of all instances.
Considering \qty{75}{\percent} of all instances, the only significant result is that \combconf{s bi} outperforms \combconf{s b} but with no advantage, i.e. $\Delta=1$.
For the smaller instances of \Cref{fig:eng-binom-ss}, the comparison looks vastly different.
Here, \combconf{s bi} outperforms all other variants on at least \qty{75}{\percent} of instances, although its advantage is not large, with only up to $1.6$ even on the most extreme quarter of the dataset.
Furthermore, variants \SPd and \combconf{s b} outperform variants \combconf{s i} and \combconf{s} on half of the dataset, but again with no noticeable advantage, that is $\Delta=1$.
To summarize, our results are again in accordance with those from \Cref{sec:eng-batch,sec:eng-combinations}, where for large instances variant \combconf{s} is the fastest, whereas for smaller instances \combconf{s bi} is superior.

\begin{figure}
  \centering
  \begin{subfigure}{.5\linewidth}
    \includegraphics[width=\linewidth]{graphics/plots/stats-clusters-large-cluster-crossing-stats_time_ns}
    \vspace{-1cm}
    \caption{\hspace*{\fill}}
  \end{subfigure}\hfill \begin{subfigure}{.5\linewidth}
    \includegraphics[width=\linewidth]{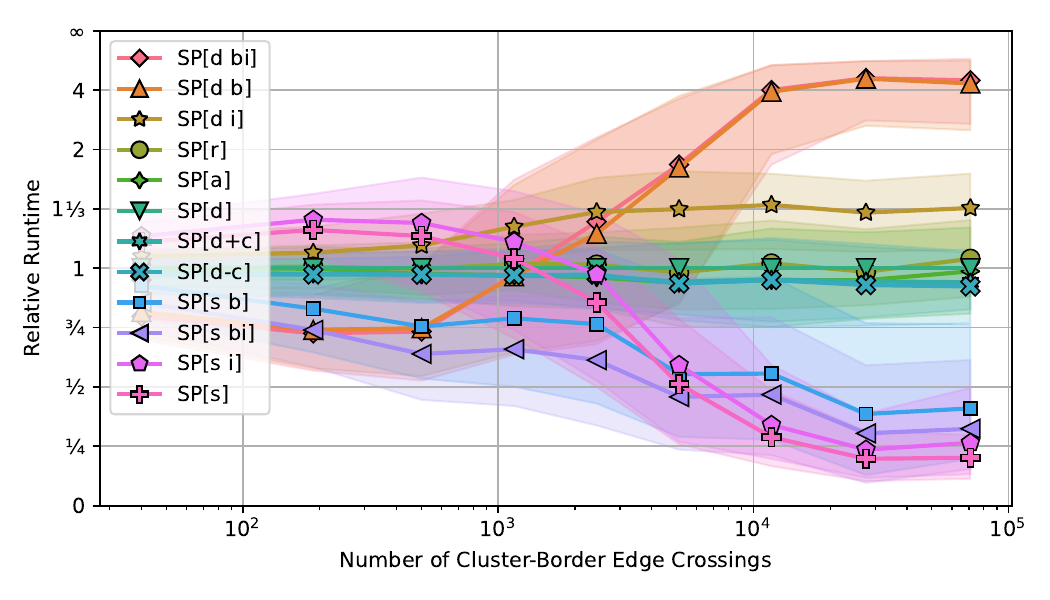}
    \vspace{-1cm}
    \caption{\hspace*{\fill}}
  \end{subfigure}
\caption{
    Absolute {(a)} and relative {(b)} running times with regard to \SPd for \dslarge.
  }
  \label{fig:eng-rt-dslarge}

  \begin{subfigure}{.5\linewidth}
    \includegraphics[width=\linewidth]{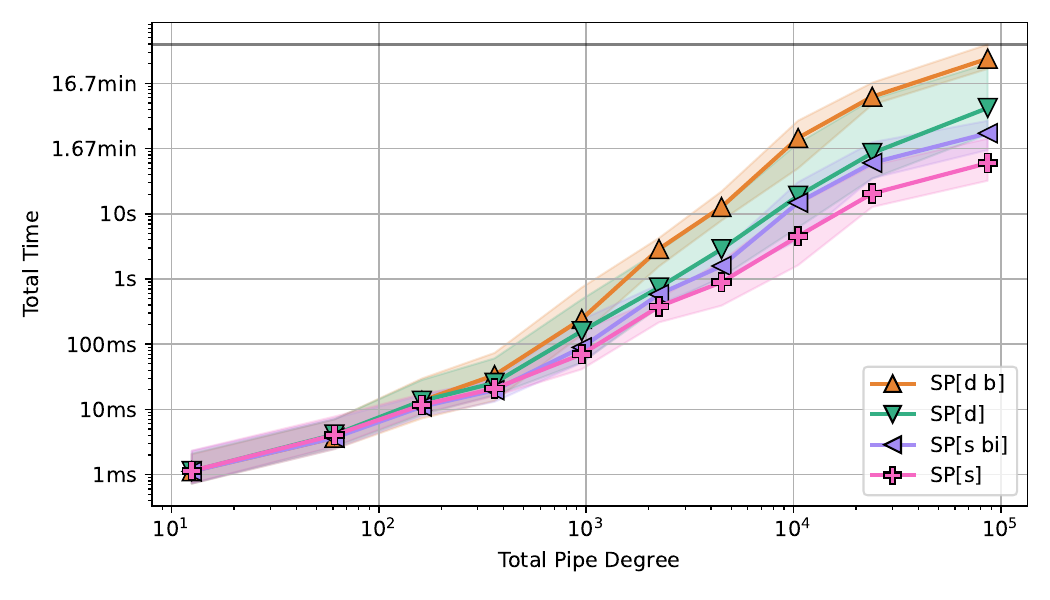}
    \vspace{-1cm}
    \caption{\hspace*{\fill}}
  \end{subfigure}\hfill \begin{subfigure}{.5\linewidth}
    \includegraphics[width=\linewidth]{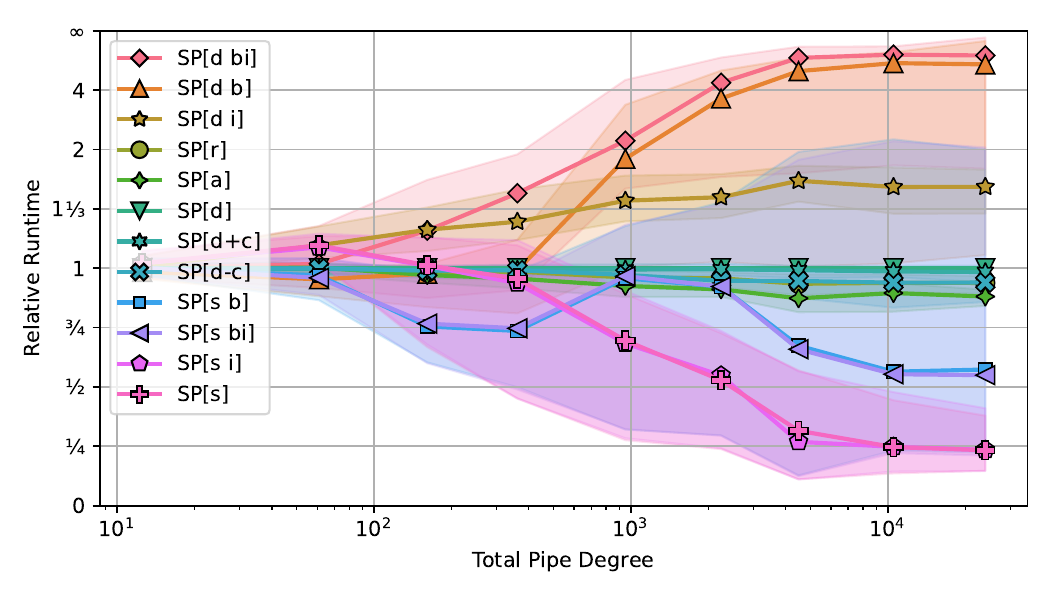}
    \vspace{-1cm}
    \caption{\hspace*{\fill}}
  \end{subfigure}
  \caption{
    Absolute {(a)} and relative {(b)} running times with regard to \SPd for \dspq.
  }
  \label{fig:eng-rt-dspq}

  \begin{subfigure}{.5\linewidth}
    \includegraphics[width=\linewidth]{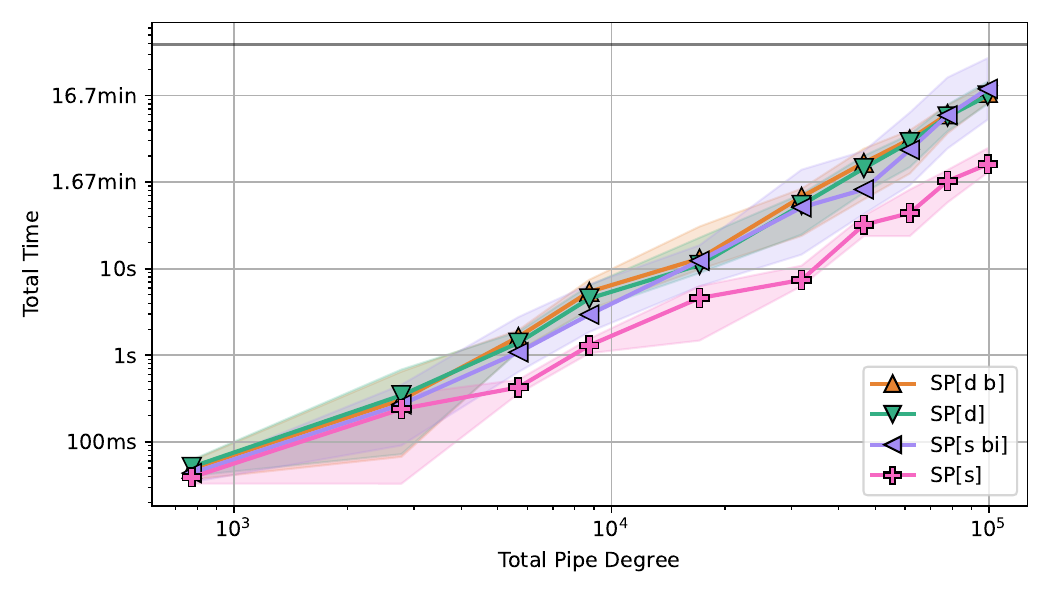}
    \vspace{-1cm}
    \caption{\hspace*{\fill}}
  \end{subfigure}\hfill \begin{subfigure}{.5\linewidth}
    \includegraphics[width=\linewidth]{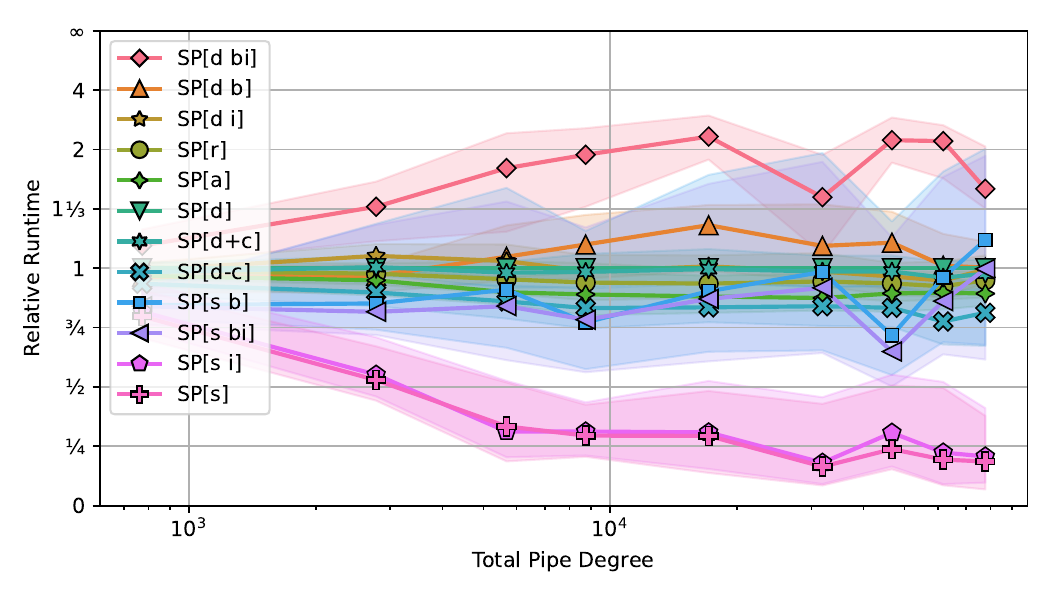}
    \vspace{-1cm}
    \caption{\hspace*{\fill}}
  \end{subfigure}
  \caption{
    Absolute {(a)} and relative {(b)} running times with regard to \SPd for \dssefe.
  }
  \label{fig:eng-rt-dssefe}
\end{figure}

\subsection{Other Problem Instances}\label{asec:other-ds}
Running the same evaluation on the datasets \dssefe and \dspq yielded absolute running times with roughly the same orders of magnitude as for \dslarge,
see the left plots in \Cref{fig:eng-rt-dslarge,fig:eng-rt-dssefe,fig:eng-rt-dspq} (but note that the plots show different ranges on the x-axis while having the same scale on the y-axis).
The right plots in the figures again detail the running times relative to \combconf{d}.
For \dspq, the relative running time behavior is similar to the behavior observed on \dslarge.
The two major differences concern variants with flag \combflag{b}.
Variant \combconf{d b(i)} is not faster than \SPd on small instances and also sooner grows slower on large instances.
Similarly, \combconf{s b(i)} is not much faster than \SPd on small instances, and its speed-up over \SPd for larger instances has a dent where it returns to having roughly the same speed as \SPd around size 1000.
On a large scale, this behavior indicates that the slowdown caused by large connected components is even worse in dataset \dspq.
For \dssefe, the instances are less evenly distributed in terms of their total pipe degree, as the total pipe degree directly corresponds to the vertex degrees in the \sefe instance.
Regarding the relative running time behavior, we still see that \combconf{d bi} is much slower and \combconf{s (i)} much faster than \SPd.
For the remaining variants, the difference to \SPd is much smaller than in the two other datasets.
This indicates that the size of connected components does not play an as important role in this dataset as before.

\section{Conclusion}
In this paper, we described the first practical implementation of \pqplan, which generalizes many constrained planarity problems such as \cplan and \consefe.
We evaluated it on more than \num{28000} instances stemming from different problems.
Using the quadratic algorithm by Bläsius et al.~\cite{bfr-spw-21}, instances with 100 vertices are solved in milliseconds, while we can still solve most instances with up to \num{100000} vertices within minutes.
This makes our implementation at least an order of magnitude faster than all other \cplan implementations, which also have a worse asymptotic running time.
Furthermore, we found incorrect results in one of the other implementations.
Analyzing our running times in more detail, we find the generation of embedding information in the form of embedding trees to be by far the most time-consuming,
while the actual operations of the algorithm that reduce and solve the instance are comparatively fast.
We apply algorithm engineering and use the various degrees of freedom of the algorithm to speed up computation times by up to an order of magnitude.
The main result here is that the batched computation of embedding information we devise using SPQR-trees produces a major speed-up.
Tuning some other variables produces a speed-up only in parts of the algorithm while slowing down others, showing that further speed-ups may be more challenging to achieve and that trade-offs may have to be made.
One possible approach could be implementing the dynamically-maintained SPQR-tree described by Fink and Rutter~\cite{fr-mtc-23}, which also yields a further theoretical speed-up.
As contribution towards future work in the field of graph drawing, we also see that our implementation can be used as reference for the implementation of more specialized, but potentially faster constrained planarity algorithms, which proved challenging in the past~\cite{brue-pvf-21}.

\bibliographystyle{abbrvurl}
\bibliography{bibliography}

\end{document}